\documentclass[fleqn,usenatbib]{mnras}

\usepackage{newtxtext,newtxmath}
\usepackage{graphicx}
\usepackage{lipsum}  
\usepackage{natbib}
\usepackage[percent]{overpic}
\usepackage[T1]{fontenc}
\usepackage{booktabs} 
\DeclareRobustCommand{\VAN}[3]{#2}
\let\VANthebibliography\thebibliography
\def\thebibliography{\DeclareRobustCommand{\VAN}[3]{##3}\VANthebibliography}

\usepackage{graphicx}	
\usepackage{amsmath}	

\def\dndz{$dN/dz$}
\def\dndb{$dN/d\beta$~}

\def\mgii{Mg~{\sc ii}}

\def\ewmgiia{W$_{r}$(2796)}

\def\civ{C~{\sc iv}~}

\def\lya{Ly$\alpha$~}

\newcommand{\kms}{\ensuremath{\text{km}\,\text{s}^{-1}}}
\defcitealias{Bergeron2011A&A...525A..51B}{BBM}
\newcommand{\ang}{\text{\AA}}

\title[Mg~{\sc ii} number density towards radio quasars] {On the incidence of weak and strong Mg~{\sc ii} absorbers towards the Flat and Steep Spectrum Radio quasars}

\author[R.Kumar et al.]{
Ritish Kumar,$^{1}$\thanks{E-mail: ritishshield@gmail.com}
Sapna Mishra,$^{2}$
Hum Chand$^{1}$\\
$^{1}$Department of Physics and Astronomical Science, Central University of Himachal Pradesh, Dharamshala, 176215, India\\
$^{2}$ Space Telescope Science Institute, 3700 San Martin Drive, Baltimore, MD 21218, USA\\
}

\date{Accepted XXX. Received YYY; in original form ZZZ}

\pubyear{\the\year{}}

\begin{document}
\label{firstpage}
\pagerange{\pageref{firstpage}--\pageref{lastpage}}
\maketitle

\begin{abstract}
The incidence rate of \mgii~absorbers per unit redshift path  (\dndz) towards quasars' sightline has been used to probe the interaction between quasar jets and surrounding gas clouds. Studies using core- and lobe-dominated samples found a higher \dndz~ for strong \mgii~absorbers (rest equivalent width, $W_{r}(2796)\geq~1.0~\ang$) in velocity offsets ranging from 5000~\kms$ <\beta c <~$60000~\kms (with $\beta \equiv v/c$), toward core-dominated sources. In this study, we applied a  stringent spectral index criterion:
$\alpha_{\text{radio}}~<-0.7$ for steep-spectrum radio quasars (SSRQs) and $\alpha_{\text{radio}} >-0.3~$ for flat-spectrum radio quasars (FSRQs). Using this, we assembled the largest sample till date --- 441 FSRQs and 464 SSRQs with suitable optical spectra --- to study both strong absorbers and weak ($0.3~\ang<W_r (2796)< 1.0$~\ang~) \mgii~absorbers. We conducted a detailed comparison of absorbers' incidence rate and offset velocity distributions.  Our main findings are: (i) For both weak and strong absorbers, we found no significant excess in \dndz~towards FSRQ compared to SSRQ sightlines. (ii) The \dndb~distribution of  \mgii~absorbers along FSRQs and SSRQs are statistically similar. (iii) The cumulative distribution of weak \mgii~absorbers is slightly lower for $\beta < 0.3$, but shows an excess at higher $\beta$. This suggests that, while intrinsic \mgii~absorber abundance is comparable along both sightlines, FSRQs’ more aligned relativistic jets cluster weak absorbers at high $\beta$, consistent with the scenario of jet-driven acceleration of smaller gas clumps.
\end{abstract}



\section{Introduction}
\label{introduction}
Quasar absorption line systems (QALs) are spectral features imprinted on a quasar’s continuum due to the presence of gas along the line of sight. These systems arise when the quasar’s continuum emission passes through foreground gas, resulting in absorption at specific wavelengths. QALs are a valuable tool for studying metal enrichment, ionization state, kinematics, and gas structure within galaxies and the intergalactic medium (IGM). They provide crucial insights into high-redshift galaxies, which are challenging to detect directly through imaging or spectroscopy, even with the largest telescopes  \citep[e.g., see][]{Bahcall1966ApJ...144..847B,steidel1992mg,Wolfe2005ARA&A..43..861W,Kulkarni2012ApJ...749..176K}. QALs can be further classified based on their velocity offset relative to the background quasars. Absorption systems with a velocity offset $\beta c < 5000$ km$~s^{-1}$ with respect to the background quasars are widely considered to be gravitationally bound to the quasars and are classified as associated absorbers \citep[e.g.,][]{Anderson1987AJ.....94..278A,berk2008average,shen2012link,chen2025detection}, whereas those with larger velocity offsets ($\beta c > 5000$ km$~s^{-1}$) are categorized as intervening absorption systems \citep[][]{bergeron1986mg,Nestor2005ApJ...628..637N,chen2023study}. Both associated and intervening absorbers are extensively studied using optical spectroscopy, primarily through strong metal absorption lines such as \mgii~$(\lambda\lambda2796, 2803)$ \cite[e.g., see][]{bergeron1991sample,Nestor2005ApJ...628..637N,Srinivasan2016MNRAS.463.2640R} and \civ~$(\lambda\lambda$1548, 1550) \cite[e.g., see][]{Anderson1987AJ.....94..278A,baker2002associated}. The intervening \mgii~absorbers are classified into two main categories based on their rest-frame equivalent width ($W_{r}$) as `strong systems'($W_{r}\geq1\ang$) and `weak systems'($0.3<W_{r}<1\ang$). The strong systems are generally associated with galaxies within an impact parameter of $\sim$100~kpc and spanning a wide range of luminosity \cite[e.g., see][]{steidel1995nature,churchill2005mgii}. On the other hand, the weak \mgii~systems are thought to predominantly trace separate populations of galaxies, such as low surface brightness galaxies or dwarf galaxies \cite[e.g., see][]{churchill1999population}.\\
It is widely established that the associated absorption system depends on the nature of the background source and is closely linked to its environment \citep[e.g., see][]{aldcroft1994mg,baker2002associated,chen2025detection}. \citet[][]{baker2002associated} analyzed the sample of 36 radio-loud quasars derived from 557 optically identified radio sources from the Molonglo Quasar Sample \citep[MQS,][]{kapahi1998molonglo}, namely steep-spectrum radio-quasars (SSRQ) and compact steep-spectrum sources (CSS). Their study found an excess in associated \civ~absorbers towards SSRQs compared to CSS, suggesting that the absorbing material tends to lie away from the radio-jet axis. On the other hand, \citet{chen2025detection} analyzed a sample of 3,141 radio-loud quasars, among which 418 exhibit \mgii~associated absorption. They classify these quasars into evolutionary stages based on their radio spectral shapes, identifying them as ``non-peaked'' (evolved) spectrum and ``gigahertz-peaked'' (young) spectrum sources, to study the dependence of \mgii~absorption on quasar evolution. Their study found that quasars with ``non-peaked'' evolved radio spectra show an incidence rate of \mgii-associated absorbers approximately 1.7 times greater than that of  ``gigahertz-peaked'' young radio sources. \par

Unlike associated absorbers, which are physically linked to the background quasar, intervening \mgii~absorbers are associated with foreground galaxies and are, therefore, expected to be completely independent of the background quasars. However, studies in the literature suggest that intervening absorption systems may not be entirely independent of the background quasars \citep[e.g., see][]{Stocke1997,Tejos2009ApJ...706.1309T,Bergeron2011A&A...525A..51B,chand2012incidence,Joshi2013MNRAS.435..346J,christensen2017solving,mishra2018incidence}. These studies estimate different incidence rates (\dndz) for intervening absorption systems observed toward various background sources, including blazars, gamma-ray bursts (GRBs), and normal QSOs. \citet[][hereafter BBM]{Bergeron2011A&A...525A..51B} analyzed the high signal-to-noise ratio (SNR) optical spectra of 42 blazars observed with Very Large Telescope / FOcal Reducer and the low-dispersion Spectrograph (VLT/FORS1) in the redshift range (0.8 $< z <$ 1.9). The authors found that the \dndz\ of strong and weak \mgii\ absorption systems in blazars is higher by factors of $\sim$2 (3$\sigma$) and $\sim$1.8 (2$\sigma$), respectively, compared to normal QSOs. They ruled out dust extinction and gravitational lensing as significant contributors to the excess in \dndz, instead attributing it to gas physically entrained by powerful blazar jets under specific conditions. On the other hand, \citet{chand2012incidence}, using high-resolution VLT/ UVES spectra of 115 flat-spectrum radio quasars (FSRQ, non-blazar type) did not find any significant excess in the \dndz~of \mgii~absorption systems as compared to normal QSOs. They attributed the lack of excess around FSRQs to their jets being less closely aligned with our line of sight compared to those of blazars. Later, \citet{Joshi2013MNRAS.435..346J} extended this study by analyzing a large, redshift-matched sample of 3975 core-dominated quasars (CDQs) and 1583 lobe-dominated quasars (LDQs), classified based on their radio spatial morphology. Their study found no significant excess in \dndz\ towards CDQs compared to normal QSOs. However, they observed a notable excess in the \dndz\ of CDQs ($3.75\sigma$) when restricting the comparison to $\beta < 0.1$ where $\beta \equiv v/c$, with $v$ representing the line-of-sight velocity and $c$ the speed of light). Recently, \citet{mishra2018incidence} analyzed spectral data for 191 blazars, three times the size of BBM's sample, but found no significant excess in overall \dndz\ for both weak and strong \mgii\ systems. Nevertheless, they did not rule out the possibility of an excess in the cumulative number of the \mgii\ absorbers towards blazars upto velocity offset of $\beta < 0.2$ relative to background QSOs. \\
These studies suggest that relativistic jets in radio-loud AGNs may influence nearby absorbers by accelerating them to relativistic speeds along the line of sight, potentially mimicking the features of intervening absorbers. However, to date, no study has specifically examined intervening absorption systems associated with FSRQs and SSRQs. The study by \citet{Joshi2013MNRAS.435..346J} analyzed radio sources classified by their spatial morphology as CDQs and LDQs, but these classifications do not represent a well-defined population of radio sources in terms of jet orientation. Motivated by the need to better understand the distribution of intervening \mgii\ absorbers across well-defined quasar subclasses with radio jets, we use a significantly larger sample of FSRQs and steep-spectrum radio quasars (SSRQs), selected using the more stringent spectral index ($\alpha_{\text{radio}}$) criteria from \citet[hereafter FR14]{Farnes2014a}, and present the corresponding \dndz\ and \dndb~distributions for these two quasar types. \par
This paper is organized as follows: In Section~\ref{sample}, we describe the selection criteria and construction of the FSRQ and SSRQ samples. Section~\ref{analysis} outlines the analysis methods applied to study the \mgii~absorber distributions.
The results of our analysis are presented in Section~\ref{results}. In Section~\ref{discussion} and Section~\ref{Sect:summary}, we discuss and summarize the key findings of the study.

\section{The Sample Selection}
\label{sample}
To construct a statistically significant sample of FSRQs and SSRQs for studying the incidence of \mgii~absorbers, we utilize a sample of 25,649 extragalactic radio sources from FR14. This catalog provides redshift information and total intensity spectral indices derived from multiwavelength linear polarization and total intensity radio data for polarized sources in the NRAO VLA Sky Survey. Next, to identify the \mgii~absorption along these radio sources, we searched for their archival spectral data. For this, we use three optical spectral archives:  Sloan Digital Sky Survey Data Release 16 \citep[SDSS-DR16;][]{wu2022catalog}, the UVES Spectral Quasar Absorption Database \citep[SQUAD;][]{murphy2019uves}, and the Keck Observatory Database of Ionized Absorption toward Quasars \citep[KODIAQ;][]{o2020third}. We cross-match the FR14 catalog with SDSS-DR16, SQUAD, and KODIAQ using a search radius of 7 arcseconds. This process identifies 1,705 sources from SDSS-DR16, 86 from SQUAD, and 25 from KODIAQ.  
To ensure sufficient spectral coverage for detecting \mgii~absorbers, we apply a redshift cut, selecting only quasars with an emission redshift of $z_{\text{emi}} > 0.4$. This criterion limits the sample to 1,641 quasars from SDSS-DR16, 62 from SQUAD, and 25 from KODIAQ. To further classify the sample into FSRQs and SSRQs, we apply the same spectral index ($\alpha_{\text{radio}}$) criteria as in FR14. Specifically, sources with $\alpha_{\text{radio}} > -0.3$ are identified as FSRQs, while those with $\alpha_{\text{radio}} < -0.7$ as SSRQs. Applying this filter reduces the sample to 1,281 sources, including 1,194 from SDSS-DR16, 62 from SQUAD, and 25 from KODIAQ. To ensure sufficient spectra data quality,  we exclude sources with a median signal-to-noise ratio (SNR) below 5, resulting in a final sample of 853 quasars. Among these, 389 sources (363 from SDSS-DR16, 21 from SQUAD, and 5 from KODIAQ) are identified as FSRQs ($\alpha_{\text{radio}} > -0.3$), while 464 sources (449 from SDSS-DR16, 9 from SQUAD, and 6 from KODIAQ) are identified as SSRQs ($\alpha_{\text{radio}} < -0.7$). \par
\begin{figure}
   {\includegraphics[width = 0.5\textwidth,height=0.35\textwidth]{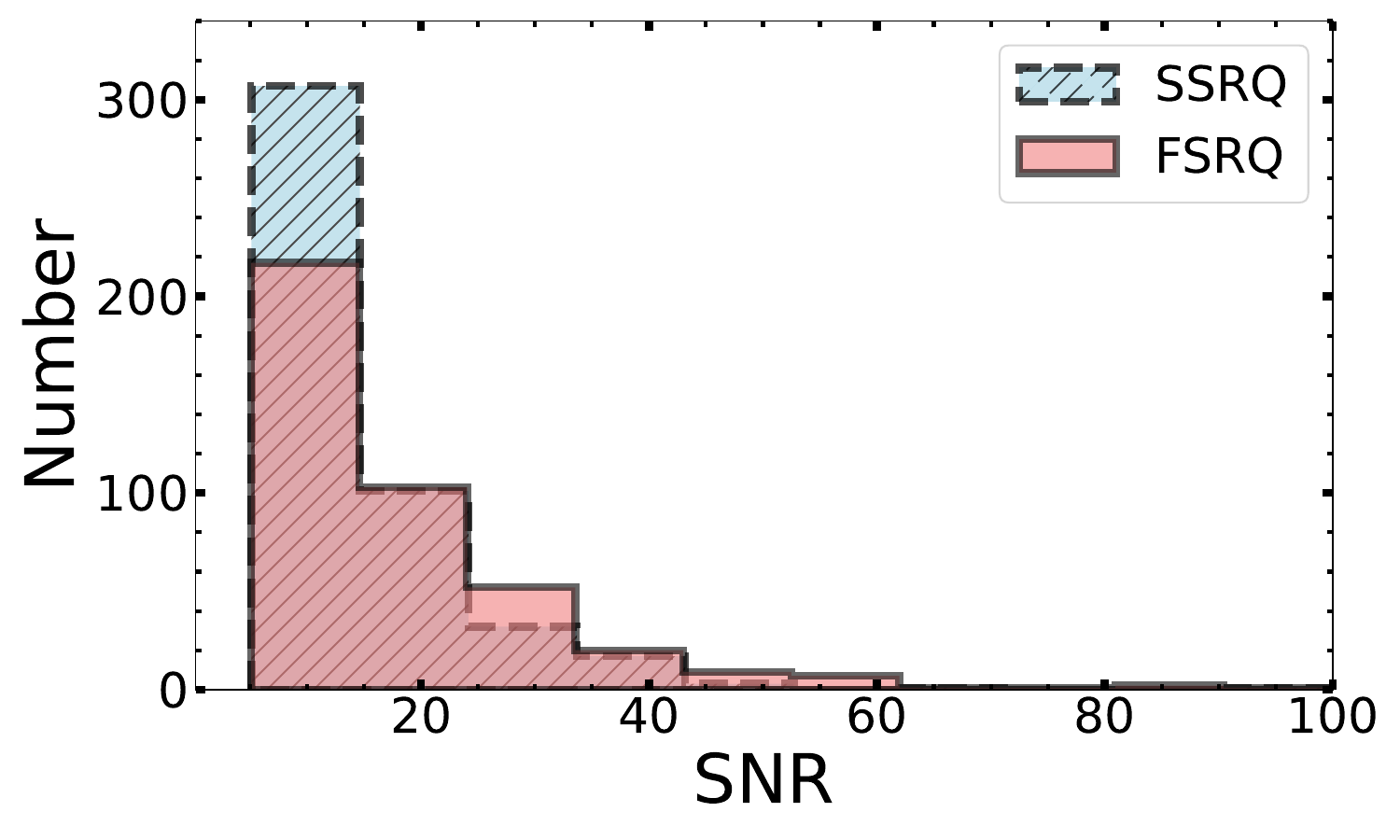}}
   \caption{Distribution of the signal-to-noise ratio (SNR) for the sample of 441 FSRQs (solid red) and 464 SSRQs (dashed steel).}
   \label{fig: snr_dist}
\end{figure}

To further expand our sample, we also searched for FSRQs among the 115 sources reported in \citet[][hereafter CG12]{chand2012incidence}, who studied the incidence of \mgii~absorbers in the high-resolution optical spectra of these FSRQs. From their sample, we identified 52 unique FSRQs with available high-resolution spectra—44 observed with UVES/VLT and 8 with HIRES/Keck—which we include in our analysis. This adds 52 FSRQs to our dataset, bringing the total to 441 FSRQs. Thus, our final sample comprises 441 FSRQs and 464 SSRQs. \\
The step-by-step procedure for compiling our sample is summarized in Table~\ref{sample_table}. The SNR distributions for our final samples of 441 and 464 FSRQs and SSRQs, respectively, are shown in Fig.~\ref{fig: snr_dist}. As detailed in Section\ref{analysis}, only 431/441 FSRQs and 443/446 SSRQs contribute to the strong \mgii\ system analysis, while 209/441 FSRQs and 125/464 SSRQs contribute to the weak \mgii\ system analysis. The full properties of all 441 FSRQs and 464 SSRQs are provided in Table~\ref{tab:fsrq_ssrq_sample}.

\begin{table}
    \setlength\tabcolsep{2pt}
    \caption{Summary of the sample selection process used to compile the FSRQ and SSRQ samples.}
    \begin{tabular}{l@{\hskip 3pt}l@{\hskip 7pt}l}
        \hline
        (1) &  Radio quasar Sample (FR14)$^{\text{a}}$  & 25,649 \\
        (2) &  SDSS-DR16 quasar sample  & 750,414 \\
        (3) &  SQUAD quasar sample  & 467 \\
        (4) &  KODIAQ quasar sample & 727 \\\\
        Step-1:& Common with FR14$^{\text{a}}$ within 7 arcsec \\[0.05in]

        (5) & (1) and (2) (FR14 and SDSS-DR16) & 1705 \\
        (6) & (1) and (3) (FR14 and SQUAD) & 86 \\
        (7) & (1) and (4) (FR14 and KODIAQ) & 25 \\\\        
        Step-2:& Redshift cut\\[0.05in] 
        
        (8) & (5) with z $>$ 0.4 (SDSS DR16) & 1641\\
        (9) & (6) and (7) with z $>$ 0.4 (SQUAD/KODIAQ) & 91\\[0.05in]
        Step-3: & Spectral index cut with & \\[0.02in]
        & (FSRQ: $\alpha_{\text{radio}} >-0.3~$, SSRQ: $\alpha_{\text{radio}} <-0.7~$) &  \\ [0.05in]
        
        (10) & (8) and (9) with SNR$>5$ and $\alpha_{\text{radio}} >-0.3~$ (FSRQs)  &389\\

        (11) & (8) and (9) with SNR$>5$ and $\alpha_{\text{radio}} <-0.7~$ (SSRQs)  &464\\\\
                
         (12) & FSRQ sample from CG12$^{\text{b}}$ &52 \\
         (13) & (10) + (12) & 441 \\
        
        \hline
        & Final Sample: FSRQ: 441, SSRQ: 464 & \\
        \hline
    \end{tabular}
    \label{sample_table}\\
    \footnotesize{$^{\text{a}}$ \citet{Farnes2014a}, $^{\text{b}}$ \cite{chand2012incidence}.}\\
\end{table}

\section{Analysis}
\label{analysis}
The spectra for our sample of 441 FSRQs and 464 SSRQs are available in reduced form in their respective archives. The spectra from KODIAQ and SQUAD are also continuum-normalized. For the SDSS spectra, we apply the continuum normalization method outlined in \citet{Mishra2022}. The continuum-normalized spectra are then used for the identification of \mgii~absorbers and estimation of the \dndz~in subsequent analyses. \par

\subsection{\mgii~Absorption Line Identification}
\label{data reduction}

\begin{figure}
   {\includegraphics[width = 0.46\textwidth,height=0.3\textwidth]{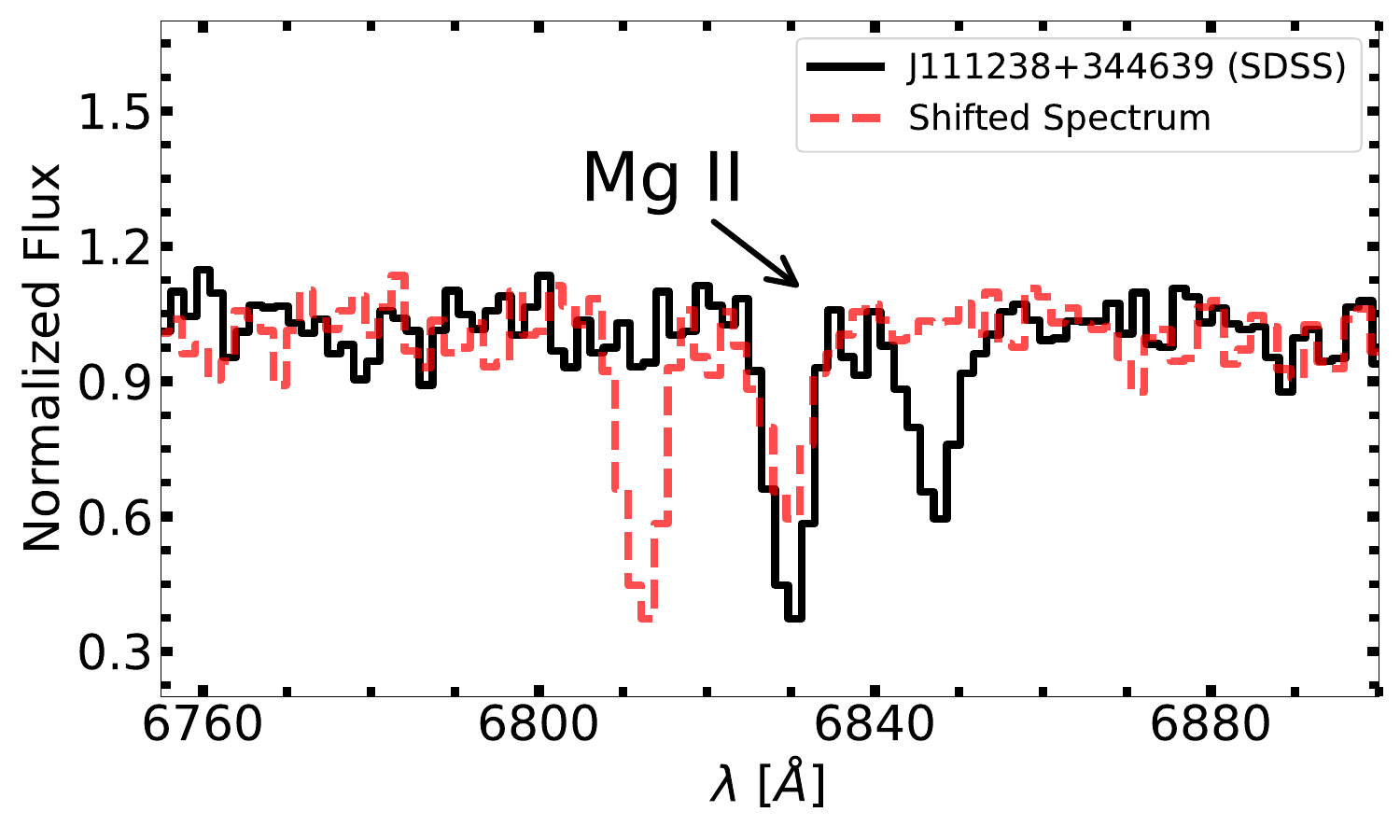}}
   \caption{The normalized SDSS spectrum of the FSRQ source J111238$+$344639 (black solid line) at an emission redshift of $z_{\text{em}} = 1.953$. An \mgii~absorption system at $z_{\text{abs}} = 1.443$ is identified by overplotting the spectrum shifted by the factor $\lambda$~2796.3543/$\lambda$~2803.5315, shown as the red dashed line.}
   \label{fig: spectrum}
\end{figure}

\begin{figure*}
   {\includegraphics[width = 0.49\textwidth,height=0.34\textwidth]{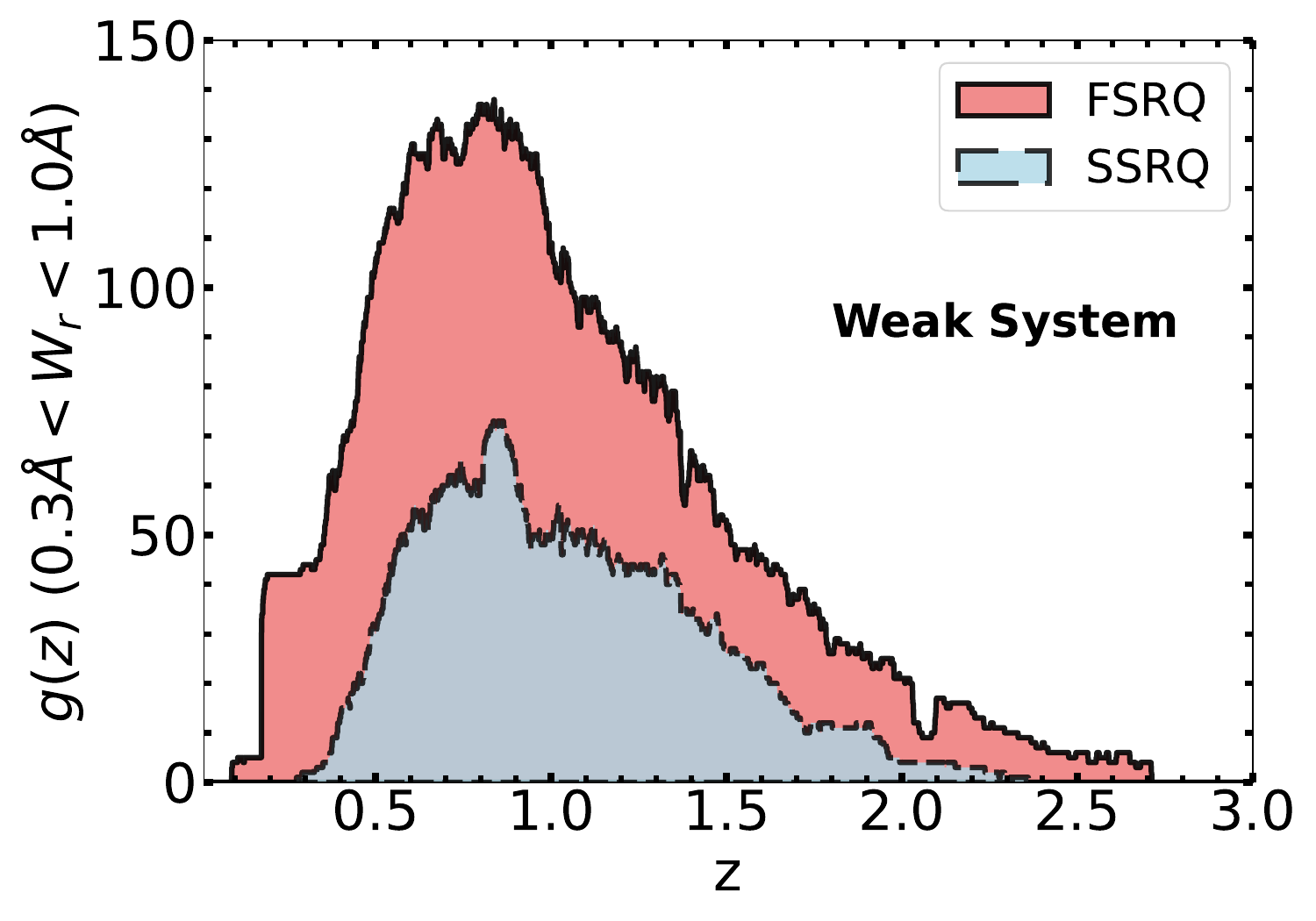}}
   {\includegraphics[width = 0.49\textwidth,height=0.34\textwidth]{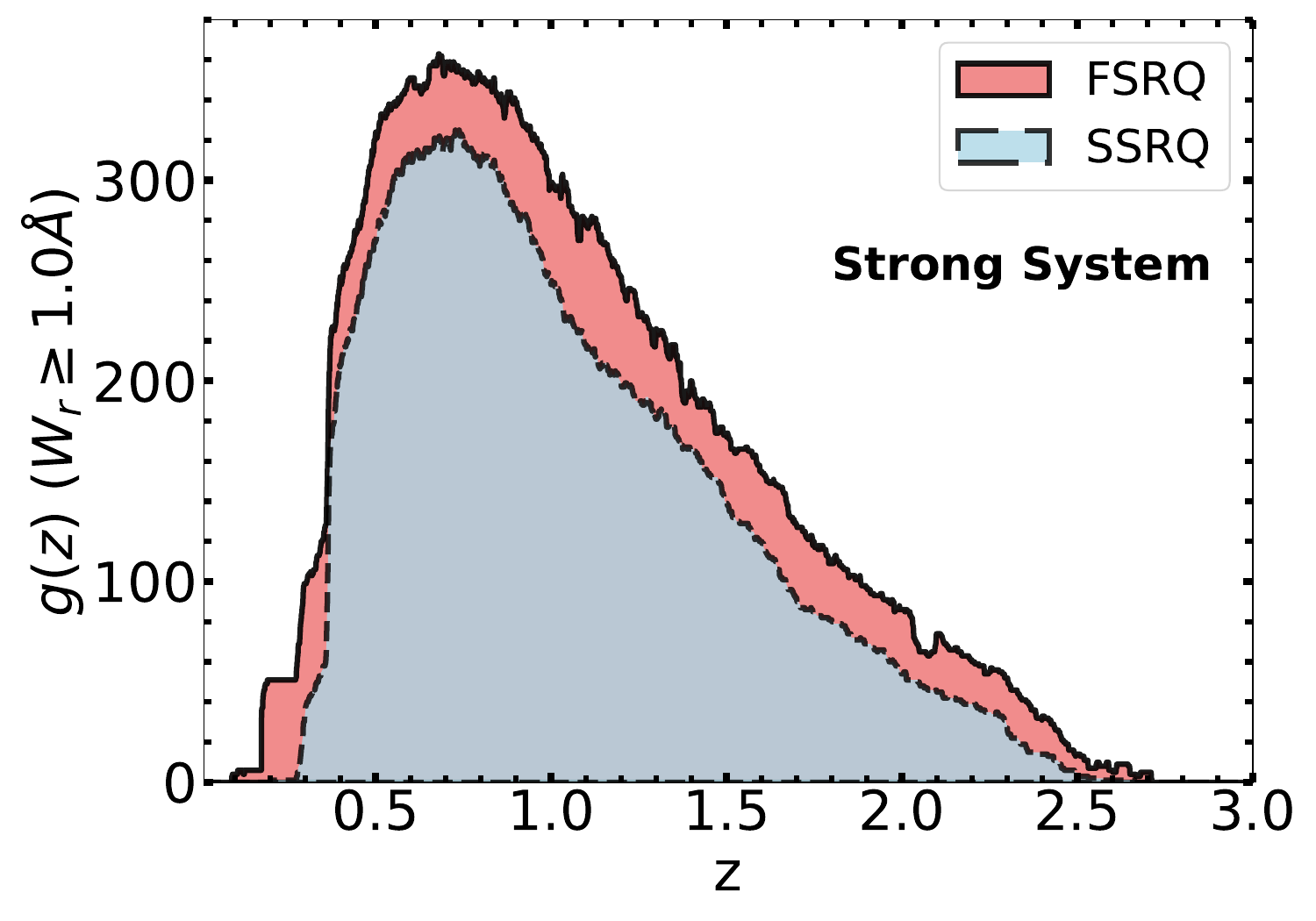}}
   \caption{\emph{Left:} Redshift path density, $g(z)$, for intervening weak \mgii\ absorbers ($0.3 < W_{r}(2796) < 1.0\text{\AA}$), excluding systems with velocity offsets below 5000~\kms$^{-1}$, shown for FSRQs (solid red-shaded region) and SSRQs (dashed blue-shaded region). \emph{Right:} Same as the left panel, but for strong  ($W_{r}(2796) \geq 1.0\text{\AA}$) \mgii\ absorption systems.}   
   \label{fig: redshift path}
\end{figure*}
To search for \mgii~absorption in our sample, we focus only on the spectral region within
$(1+z_{\text{em}}) \times 1216$~\ang~ $< \lambda < (1+z_{\text{em}}) \times 2803$~\ang. 
The lower limit is chosen to avoid contamination from the \lya forest, whereas the upper limit is set by \mgii~emission line of quasars to exclude the associated systems to the background quasars. Subsequently, we apply a profile-matching methodology similar to that of \citet{mishra2018incidence}. In brief, for each quasar in our sample, we plot the normalized spectrum and then overlay the same spectrum with the wavelength axis shifted by a factor of \mgii~$\lambda2796.3543$/\mgii~$\lambda2803.5315$ (i.e., 0.997). Next, we manually inspect spectral segments approximately 50~\ang~ wide. The regions where the 2803~\ang~absorption line in the shifted spectra aligns perfectly with the 2796~\ang~line in the unshifted spectra are identified as detected \mgii~ absorption systems. Additionally, we also inspect for the matching profile shapes and doublet ratios between the two \mgii~doublet lines \citep[see also][]{chand2012incidence}. To avoid any association of \mgii~absorbers with the background QSOs, we excluded the region within 5000 \kms of the \mgii~$\lambda\lambda$ 2796~\ang, 2803~\ang~emission line at the quasar's redshift. In Fig.~\ref{fig: spectrum}, we illustrate an example of the \mgii~identification in the SDSS spectrum of an FSRQ named  J111238$+$344639 from our sample. In total, we visually identify 241 (102) \mgii~absorbers with \ewmgiia $>$ 0.3~\ang\ (\ewmgiia $>$ 1.0~\ang) in FSRQs and 207 (109) \mgii~absorbers with \ewmgiia $>$ 0.3~\ang\ (\ewmgiia $>$ 1.0~\ang) in SSRQs.\par

\subsection{Redshift path estimation}
\label{redshift_path_estimation}
To estimate the useful spectral redshift path along each quasar sightline, we define a pixel-by-pixel sensitivity in each spectrum. The spectral sensitivity is determined following the method outlined in \citet{mishra2018incidence}. Briefly, for each pixel in the spectrum, we first compute the absolute value of \( (1 - \) normalized flux\(), \) defining it as the spectral amplitude. This amplitude is then smoothed using a top-hat filter over approximately 2–6 pixels \citep[see][]{Zhu2013ApJ...770..130Z}. Next, we identify spectral regions where the amplitude exceeds \( 1\sigma \), clip these regions, and linearly interpolate across them to remove strong spectral features (i.e., emission and absorption). In the resulting smoothed amplitude spectrum, the amplitude at a given spectral (i.e., wavelength) pixel \( i \) represents the noise, \( n_i \), for that pixel. We then estimate the \( 3\sigma \) detection threshold by calculating the limiting equivalent width (\( W_{i,\text{lim}} \)), assuming a Gaussian profile with a full width at half maximum (FWHM) of 2–6 pixels. The Gaussian amplitude is set to three times the value of the amplitude spectrum at that pixel. A spectral pixel is considered part of the `useful' redshift path if its limiting equivalent width (\( W_{i,\text{lim}} \)) is below the threshold EW (\( W_{\text{th}} = 0.3\)~\ang\ for weak and 1.0~\ang\ for strong \mgii\ systems). Otherwise, the pixel is excluded from the useful redshift path distribution. \par
Consequently, the total usable redshift path for quasars in our sample at a given redshift $z_{i}$ for detecting the \mgii\ doublet above a specified rest-frame equivalent width threshold, $W_{\rm th}$, is given by: 
\begin{eqnarray} 
\begin{aligned} 
g\left(W_{\rm th},z_{i} \right) & = \sum_{j=1}^{N_{\rm quasar}}\ H(z_i - z_{j, \rm min})\times H(z_{j, \rm max} - z_i )\\ & \times H(W_{\rm th} -{W_{j,\rm lim}(z_{i})}) \\ 
\end{aligned} 
\label{eq:goz1} 
\end{eqnarray}
where $H$ represents the Heaviside step function, and the summation is over all quasar spectra in our samples (i.e., FSRQs and SSRQs). The parameters $z_{j,\rm min}$ and $z_{j,\rm max}$ correspond to the minimum and maximum absorption redshift limits allowed for the \mgii~doublet search for the $j^{th}$ quasar. The threshold rest-frame equivalent width, $W_{\rm th}$, is set at 1.0~\ang\ and  0.3~\ang\ for the strong and weak system analysis in this study. The term $W_{j, \rm lim}(z_{i})$ denotes the estimated rest-frame limiting equivalent width at the $i^{th}$ pixel in the spectrum of the $j^{th}$ quasar. In Fig.~\ref{fig: redshift path}, we show the distribution of $g(W_{\rm th},z)$ as a function of redshift for weak ($W_{\rm th}$ = 0.3~\ang, left panel) and strong ($W_{\rm th}$ = 1.0~\ang, right panel) \mgii\ system analysis for our FSRQ and SSRQ samples.\par
 
\subsection{Estimation of dN/dz and dN/d$\beta$}
\label{beta_path_estimation}
The incidence rate of \mgii~absorbers per unit redshift path interval at a given EW threshold is expressed as \dndz$ = N_{\text{obs}}/\Delta z$; where N$_{\text{obs}}$ denotes the total count of \mgii~absorbers at a given EW threshold and  $\Delta z$ is the full effective redshift path ($\Delta$z) for a sample (e.g., FSRQs and SSRQs), is defined as:
\begin{equation}
    \Delta z=\int_{0}^{\infty} \sum_{j=1} g_{j}(W_{\rm min},z_{i})dz_{i}
    \label{eq.2}
\end{equation}
where $g_{j}(W_{\rm min},z_{i}) = 1$, if $W_{\rm th}$ ($W_{th}~=~$0.3~\ang~for weak and 1~\ang~for strong systems) is more than the $3\sigma$ limiting EW (see Section~\ref{redshift_path_estimation}), i.e., $W_{j,\rm lim(z_{i})}$ estimated for the $i^{th}$ spectral pixel; otherwise $g(W_{\rm min},z_i) = 0$. 
In our parent sample of 441 FSRQs and 464 SSRQs, a non-zero $g(W_{\rm min},z_i)$ was found for only 431 FSRQs and 443 SSRQs for the strong absorbers case, and 209 FSRQs and 125 SSRQs for the weak absorption system and are plotted in Fig.~\ref{fig: redshift path}. Table~\ref{tab:fsrq_ssrq_sample} summarizes the $\Delta$z values for weak ($0.3 < W_{r}(2796) < 1.0~\text{\AA}$) and strong ($W_{r}(2796) \geq 1.0~\text{\AA}$) \mgii~absorbers, along with their counts across individual sightlines in our full sample of FSRQs and SSRQs. The uncertainties in the calculated \dndz~values were derived using Poisson small-number statistics, ensuring a 1$\sigma$ confidence level within a Gaussian framework, as outlined by \cite{Gehrels1986ApJ...303..336G}. \par
To further investigate the role of jets in our quasar samples (FSRQs and SSRQs), we analyze the incidence rate of intervening \mgii~absorbers as a function of the relative velocity between the quasars and absorbers, quantified by a $\beta$ parameter defined as:
\begin{equation}
    \beta=\frac{v}{c}=\frac{(1+z_{\text{em}})^2-(1+z_{\text{abs}})^2}{(1+z_{\text{em}})^2+(1+z_{\text{abs}})^2}
    \label{eq:beta}
\end{equation}
where $z_{\text{em}}$ and $z_{\text{abs}}$ are the redshift of the background quasar and the \mgii~absorption system, respectively. We estimated dN/d$\beta$ as $ = N_{\text{obs}}/\Delta \beta$ with $\Delta \beta$ determined in a manner similar to $\Delta z$ (see Section~\ref{redshift_path_estimation}). The $\beta$ values for each valid pixel are calculated relative to the quasar's emission redshift using Equation~\ref{eq:beta}.

\begin{table}
    \centering
    \setlength\tabcolsep{2pt}
    \caption{Basic parameters of the 441 FSRQs and 464 SSRQs in our sample.}
    \begin{tabular}{lccrcrclc} 
        \hline
        Name & $z_{\text{em}}$ & $\alpha_r$ & $\Delta z_{\rm s}$ & N$_{s}$ & $\Delta z_{\rm w}$ & N$_{w}$ & Cat. & Class. \\
        (1) & (2) & (3) & (4) & (5) & (6) & (7) & (8) & (9)\\
        \hline
        J002903$+$050934 & 1.633  & $-$0.15 & 1.539 & 3 & 0.390 & 0 & SDSS & FSRQ \\
        J080447$+$101523 & 1.956  & $-$0.98 & 0.205 & 0 & 0.831 & 0 & KECK & SSRQ \\
        J162031$+$490153 & 1.513  & $-$0.07 & 1.901 & 1 & 0.718 & 0 & SDSS & FSRQ \\
        J215324$+$053618 & 1.967  & $-$0.88 & 3.088 & 0 & 1.544 & 0 & UVES & SSRQ \\
        J232733$+$094009 & 1.843  & $-$0.07 & 2.375 & 1 & 0.868 & 0 & SDSS & FSRQ \\
        --- & --- & --- & --- & --- & --- & --- & --- & --- \\
        \hline
    \end{tabular}
    \vspace{1mm}
    \begin{minipage}{\columnwidth}
    \textbf{Note:} The entire table is available in an online version. Only a portion of this table is shown here to display its form and content. (1) Source name. (2)–(3) Redshift and radio spectral index ($\alpha_{\text{radio}}$) for FSRQs and SSRQs. (4) and (5) Redshift path and Number of absorbers for strong system analysis (W${r}$(2796) $\geq$ 1.0 \ang). (6) and (7) Redshift path and Number of absorbers for weak system analysis 0.3 < W${r}$(2796)< 1.0 \ang. (8) Catalog corresponding to each source. (9) Source classification.
    \end{minipage}
    \label{tab:fsrq_ssrq_sample}
\end{table}

\section{Results}
\label{results}

\subsection{Redshift distribution of intervening \mgii~absorbers in FSRQs and SSRQs}
\label{subsec:result_dndz}
\begin{figure}
   {\includegraphics[width =0.45\textwidth,height=0.32\textwidth]{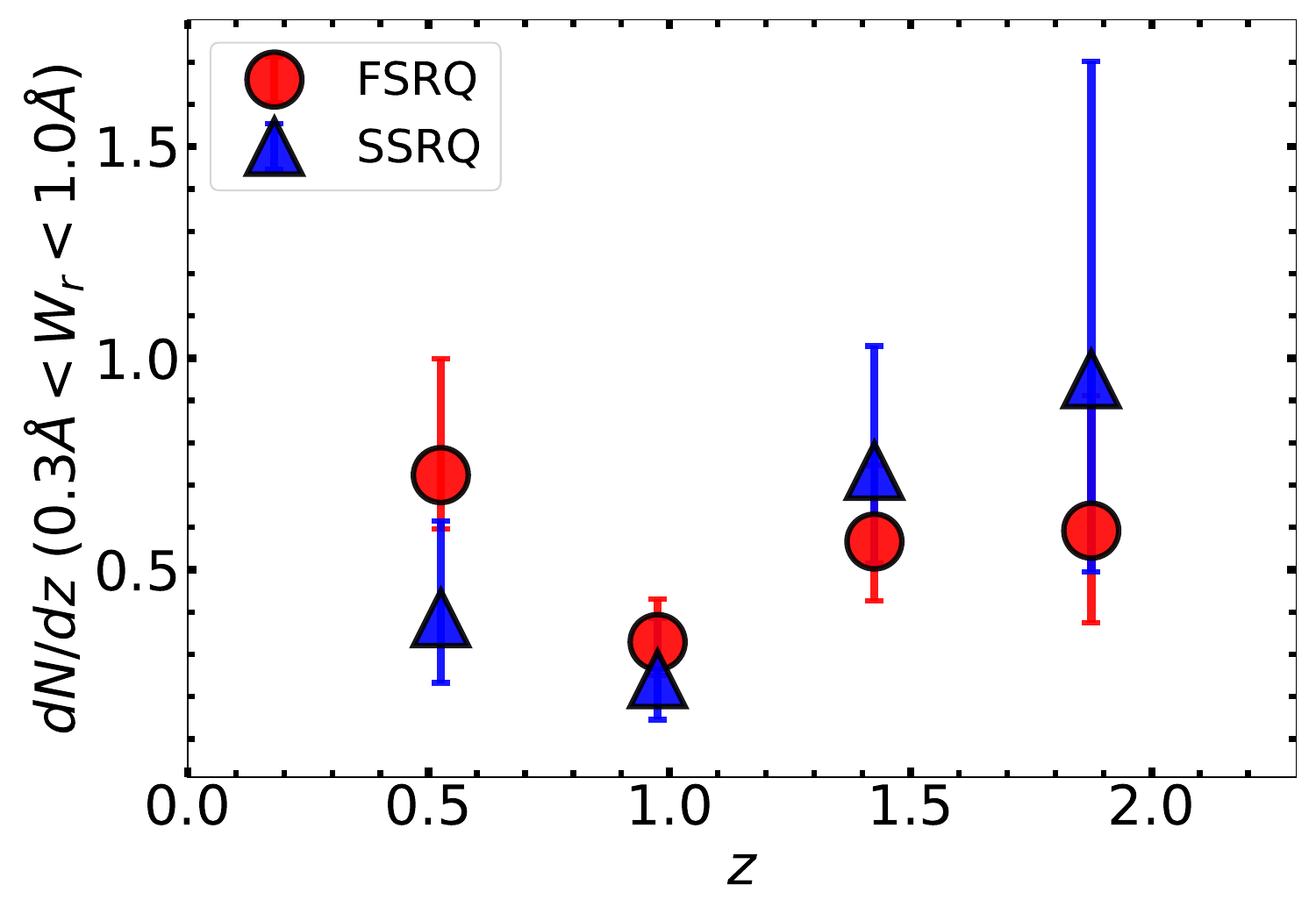}}
   {\includegraphics[width = 0.47\textwidth,height=0.33\textwidth]{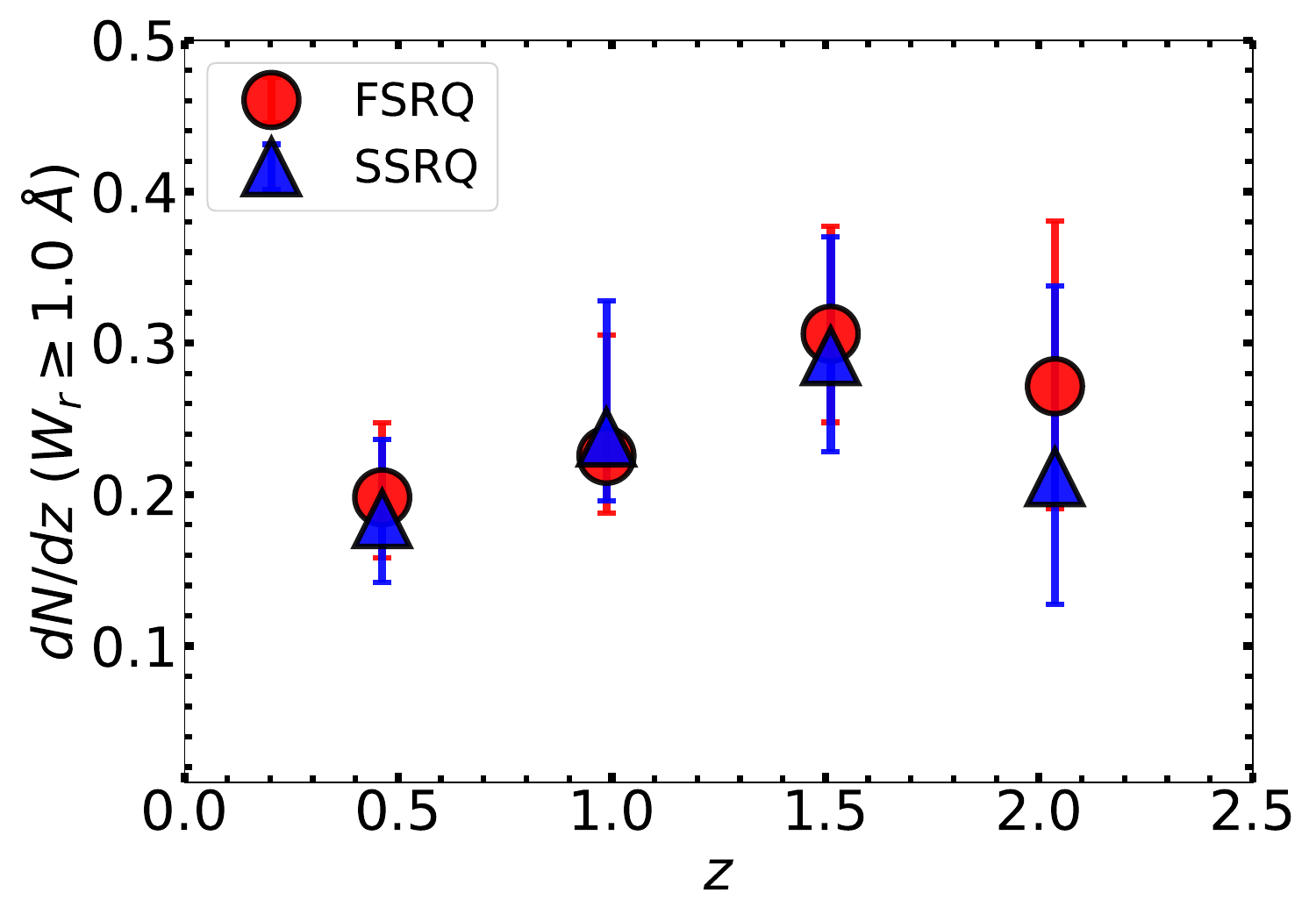}}
   \caption{ Number density evolution of weak ($0.3~\ang < W_r< 1.0~\ang$, top panel) and strong ($W_r\geq1.0~\ang$, bottom panel) \mgii~absorption systems, averaged over a redshift bin of 0.7,  towards FSRQs (red circles) and SSRQs (blue triangles) samples. The incidence rate, \dndz, corresponding to FSRQs and SSRQs, follows a similar distribution, with KS-test probability; P$_{null}$ of 0.9 for weak systems and 0.99 for strong systems, respectively.}
   \label{fig: dn_dz}
\end{figure}

\begin{figure*}
   {\includegraphics[width = 0.49\textwidth,height=0.35\textwidth]{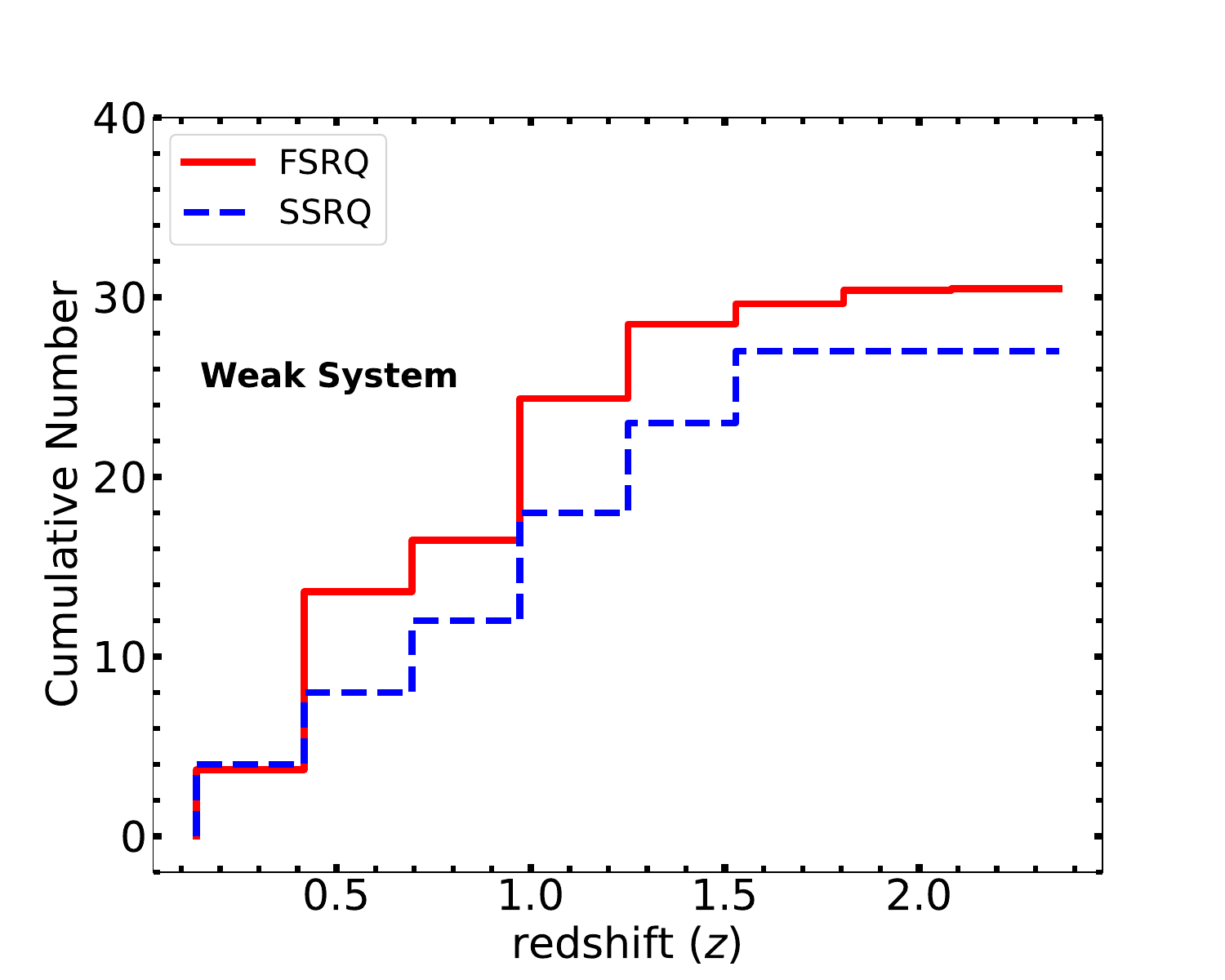}}
   {\includegraphics[width = 0.49\textwidth,height=0.35\textwidth]{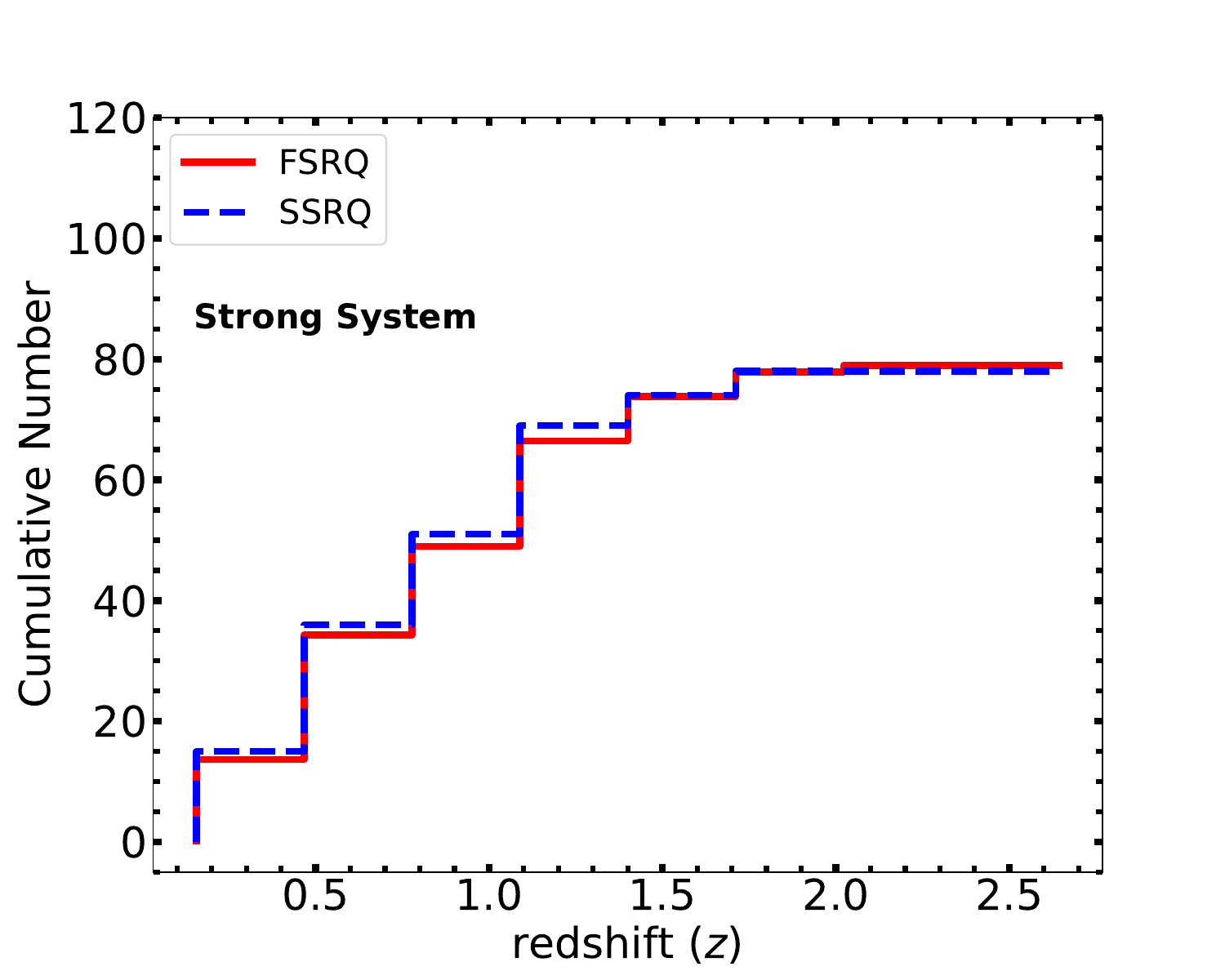}}
   \caption{Cumulative number for weak ($0.3~\ang<W_{r}(2796)<1~\ang$) (left panel) and strong ($W_{r}(2796)\geq1~\ang$) (right panel) \mgii~absorption systems detected towards the FSRQs (solid red-line) and SSRQs (dashed blue-line) after excluding the systems with a velocity offset of $5000$~\kms. To ensure a comparable redshift path length between FSRQ and SSRQ sightlines, we scaled the number of \mgii~absorbers along FSRQs within each redshift bin by the ratio of the redshift path length ($\Delta z$) of FSRQs to that of SSRQs (see text in Section~\ref{subsec:result_dndz}).}
   \label{fig: cummulative_dz}
\end{figure*}
In this section, we compare the incidence rate and the cumulative number of the \mgii~absorbers  as a function of redshift for weak and strong intervening \mgii~systems along FSRQ and SSRQ sightlines. In Fig.~\ref{fig: dn_dz}, we show the redshift evolution of \dndz\ for \mgii\ absorbers along FSRQs and SSRQs, averaged over redshift bins of width 0.7 and plotted separately for weak ($0.3\ang < W_{r}(2796) < 1.0\ang$; top panel) and strong ($W_{r}(2796) \geq 1.0\ang$; bottom panel) systems. As shown, the \dndz\ distributions follow similar trends for both FSRQs and SSRQs. This similarity is supported by two-sample Kolmogorov-Smirnov (KS) test results, which yield P$_{\text{null}}$\footnote{P$_{\text{null}}$ represents the probability that the two distributions are drawn from the same parent population.} values of 0.90 and 0.99 for the weak and strong absorber samples, respectively, indicating no statistically significant difference. \\
Table~\ref{dndz_result_table} summarizes the detailed \dndz\ measurements. Our full FSRQ and SSRQ samples are compiled from multiple archival datasets with varying spectral resolutions (see Section~\ref{sample}). To evaluate whether these differences affect the comparison of \mgii\ absorber incidence rates between the two quasar types, we perform the analysis on three subsamples: the Full Sample, SDSS-only, and Non-SDSS data. As shown in Table~\ref{dndz_result_table}, the results consistently indicate no significant excess in the average \dndz\ for strong and weak \mgii\ absorbers toward FSRQs or SSRQs across various subsamples (see Column 11). \par
Further in Fig.~\ref{fig: cummulative_dz}, we present the cumulative number of \mgii\ absorbers as a function of absorption redshift for weak (left panel) and strong (right panel) systems, excluding absorbers with velocity offsets $\Delta V < 5000$ \kms. These cumulative distributions are computed over matched redshift intervals for FSRQ and SSRQ sightlines. This is because we only include absorbers within the useful redshift path, which can vary between the two quasar populations and between the analyses of weak and strong \mgii\ systems. This is clearly evident in the left panel of Fig.~\ref{fig: redshift path}, where SSRQs exhibit a significantly smaller usable redshift path than FSRQs for the weak \mgii\ systems. Therefore, to account for the variation in the redshift path lengths between the FSRQ and SSRQ samples, we normalize the number of \mgii~absorbers for FSRQs within each redshift bin by a factor, 
$\Delta z_{\text{FSRQ}}/\Delta z_{\text{SSRQ}}$ where $\Delta z_{\text{FSRQ}}$ and $\Delta z_{\text{SSRQ}}$ represent the redshift paths of FSRQs and SSRQs, respectively, in that bin. This normalization ensures comparable redshift path lengths for both samples, allowing a fair comparison of the cumulative number of \mgii~absorbers across redshift bins. In the left panel of Fig.~\ref{fig: cummulative_dz}, we observe a mild excess (P$_{\text{null}} = 0.35$) of the weak normalized cumulative number of \mgii~absorbers along FSRQs compared to SSRQs at all redshifts. We do not find such excess in the normalized cumulative number of strong \mgii\ absorbers for FSRQs and SSRQs, as evident in the right panel of Fig.~\ref{fig: cummulative_dz}.\par

\begin{table*}
\caption{Comparison of the incidence of \mgii\  absorbers towards the FSRQ and SSRQ samples.}
\centering  
\setlength{\tabcolsep}{4pt}  
\begin{tabular}{clccccccccccc}
  \toprule
  \toprule
                               & \multicolumn{3}{r}{FSRQ} & \multicolumn{4}{r}{SSRQ} & \multicolumn{3}{r}{Excess} \\ [0.1in]
  Sample & System & $N_{\text{total}}$ & $N_{\text{obs}}$ & $\Delta$z & $\frac{N_{\text{obs}}}{\Delta z}$ & $N_{\text{total}}$  & $N_{\text{obs}}$ & $\Delta$z & $\frac{N_{\text{obs}}}{\Delta z}$ &          $\frac{\left( \frac{dN}{dz} \right)_{\text{FSRQ}}}{\left( \frac{dN}{dz} \right)_{\text{SSRQ}}}$ \\[0.1in]
  (1) & (2) & (3) & (4) & (5) & (6) & (7)  & (8) & (9) & (10) & (11) \\[0.1in]
  \midrule[0.01in] 
Full Sample &  Weak & 209  & 75 & 146.90 & $0.51_{-0.07}^{+0.07}$ & 125 & 27 & 60.30 & $0.45_{-0.1}^{+0.1}$ & $1.13_{-0.29}^{+0.29}$ \\[0.15in]
                     & Strong & 431 & 97 & 414.16 & $0.23_{-0.02}^{+0.02}$& 443 & 78 & 341.8 & $0.23_{-0.03}^{+0.03}$ & $1.0_{-0.16}^{+0.16}$ \\ [0.15in]
    \midrule [0.01in]
SDSS & Weak & 136 & 36 & 79.01 & $0.45_{-0.07}^{+0.08}$ & 118 & 23 & 56.77 & $0.41_{-0.08}^{+0.1}$ & $1.10_{-0.29}^{+0.26}$ \\[0.1in]
                     & Strong & 358 & 76 & 346.248 & $0.22_{-0.02}^{+0.02}$& 436 & 78 & 337.83 & $0.23_{-0.02}^{+0.02}$ & $0.96_{-0.13}^{+0.13}$ \\ [0.15in]
    \midrule [0.01in]
Non-SDSS & Weak & 73  & 39 & 67.89 & $0.57_{-0.09}^{+0.1}$& 7& 4 & 3.52 & $1.14_{-0.5}^{+0.9}$ & $0.5_{-0.40}^{+0.24}$ \\[0.1in]
                     & Strong & 73  & 21 & 67.92 & $0.31_{-0.07}^{+0.08}$& 7 & 0 & 3.95 & --- & --- \\
  \bottomrule
\end{tabular}
\vspace{0.1in} 
\\
\parbox{0.85\textwidth}{\textbf{Note:}(1) Sample type. (2) System type: weak = $0.3~\ang < W_{r}(2796) <1.0~\ang$, strong: $W_{r}(2796) \ge 1.0~\ang $. (3) Number of FSRQs contributing to the useful redshift path analysis. (4) Number of \mgii~absorbers detected along FSRQ sightlines. (5) Total redshift path along FSRQs. (6) Average incidence rate of \mgii~absorbers along FSRQs. (7) Number of SSRQs contributing to the useful redshift analysis. (8) Number of \mgii~absorbers detected along SSRQ sightlines. (9) Total redshift path along SSRQs. (10) Average incidence rate of \mgii~absorbers along SSRQs. (11) Excess incidence rate in FSRQs compared to SSRQs.}
\label{dndz_result_table}
\end{table*}


\subsection{\texorpdfstring{$\beta$ distribution of intervening Mg{\sc ii}~ absorbers in FSRQs and SSRQs}{β distribution of intervening Mg{\sc ii} absorbers in FSRQs and SSRQs}}

\label{subsec:result_dndb}
\begin{figure}
   {\includegraphics[width=0.48\textwidth, height=0.34\textwidth]{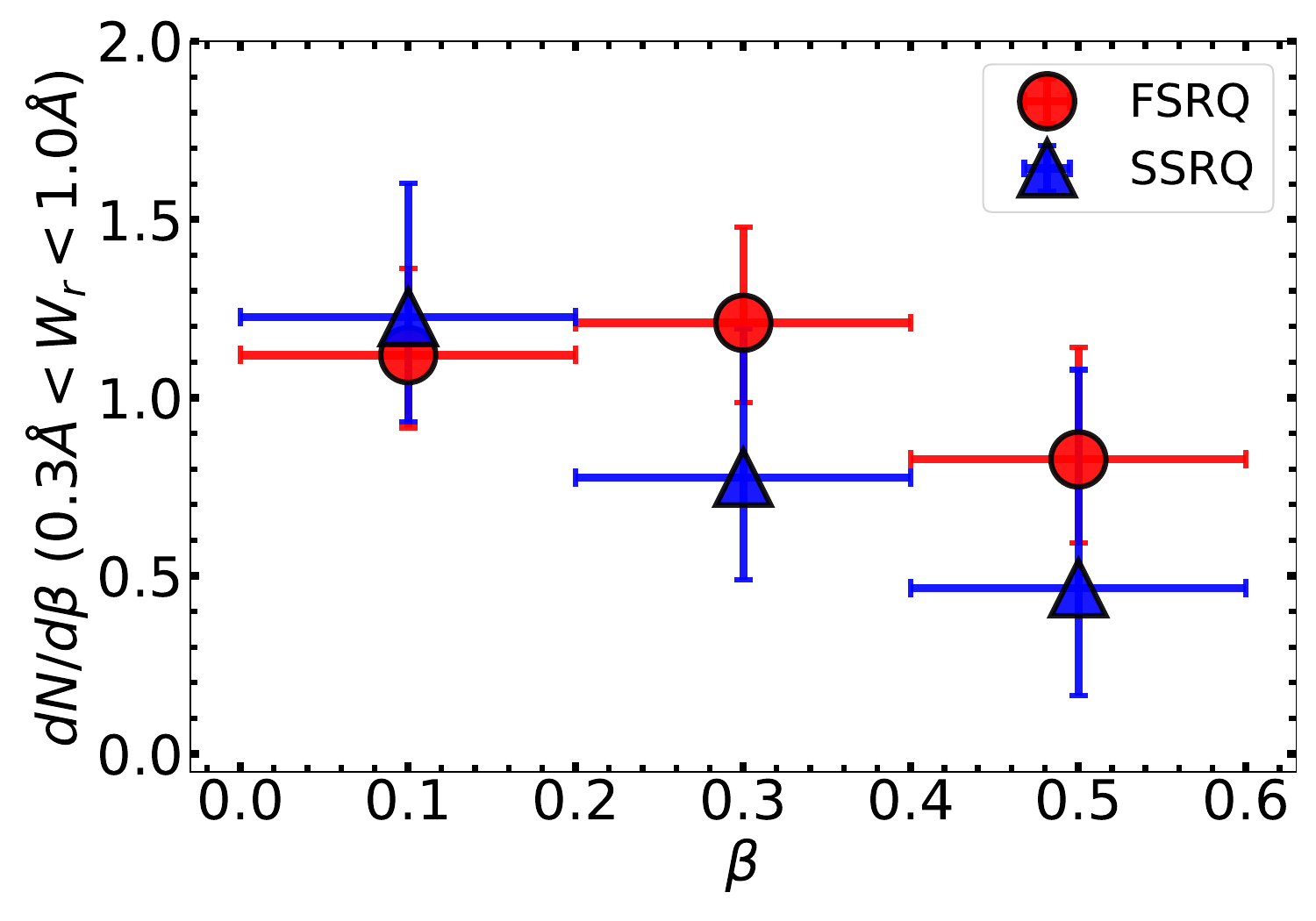}}
   {\includegraphics[width=0.48\textwidth, height=0.34\textwidth]{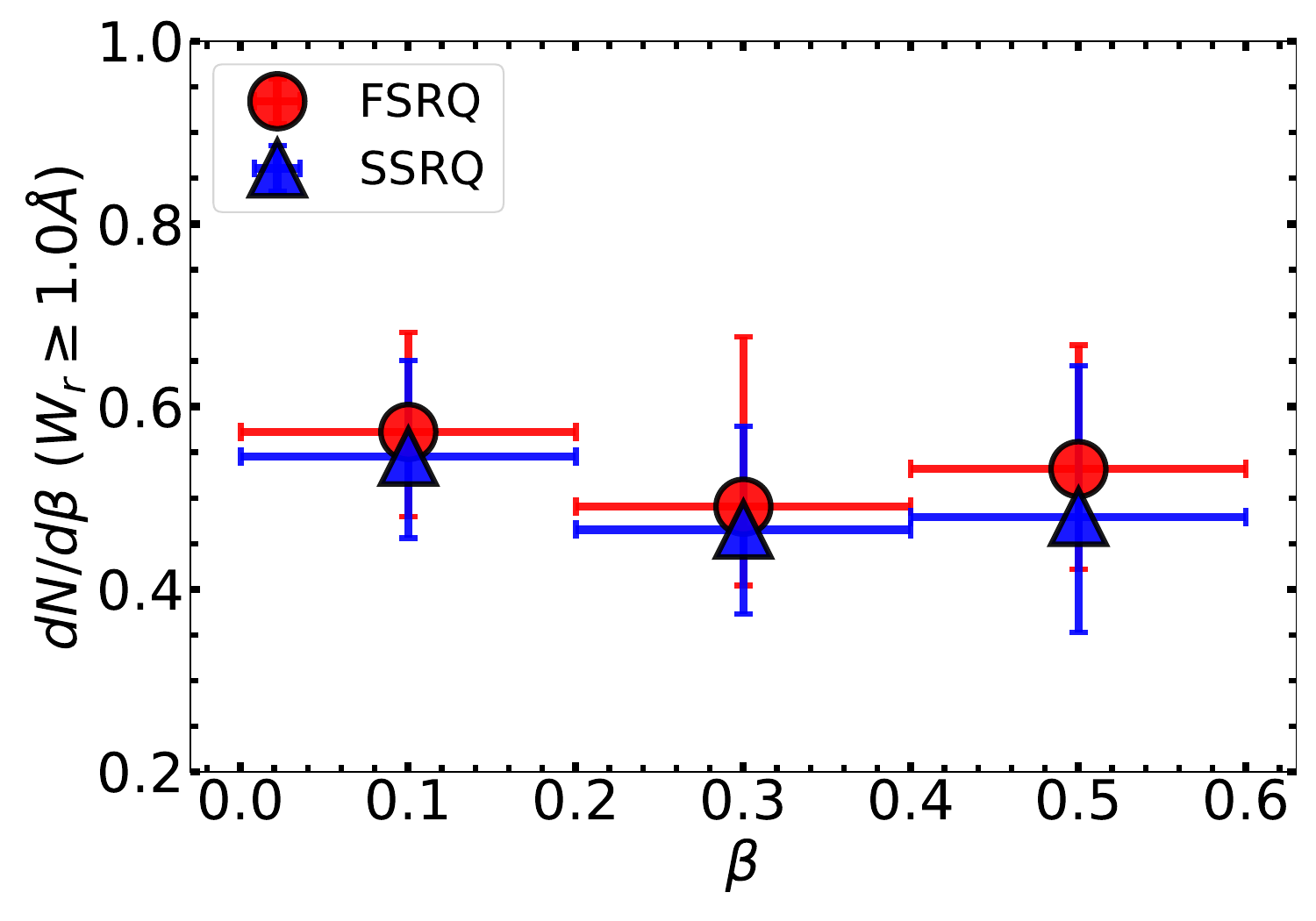}}
   \caption{ The incidence of weak (top panel) and strong (bottom panel) \mgii\ absorbers as a function of offset velocity, v = $\beta$c, measured relative to the quasar rest frame, for FSRQs (red circles) and SSRQs (blue triangles).}
   \label{fig: dn_db}
\end{figure}

\begin{figure}
   {\includegraphics[width =0.49\textwidth,height=0.35\textwidth]{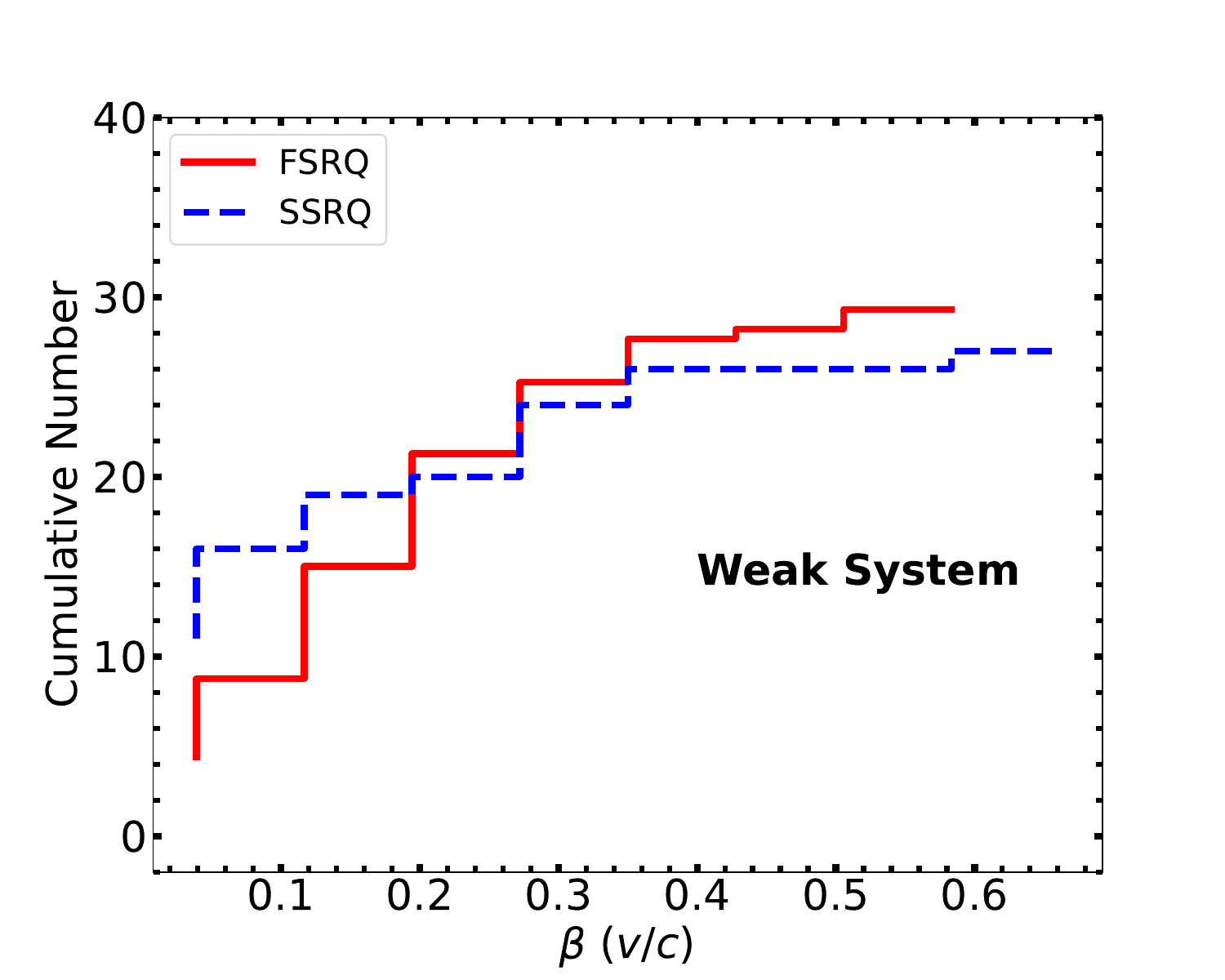}}
   {\includegraphics[width =0.49\textwidth,height=0.35\textwidth]{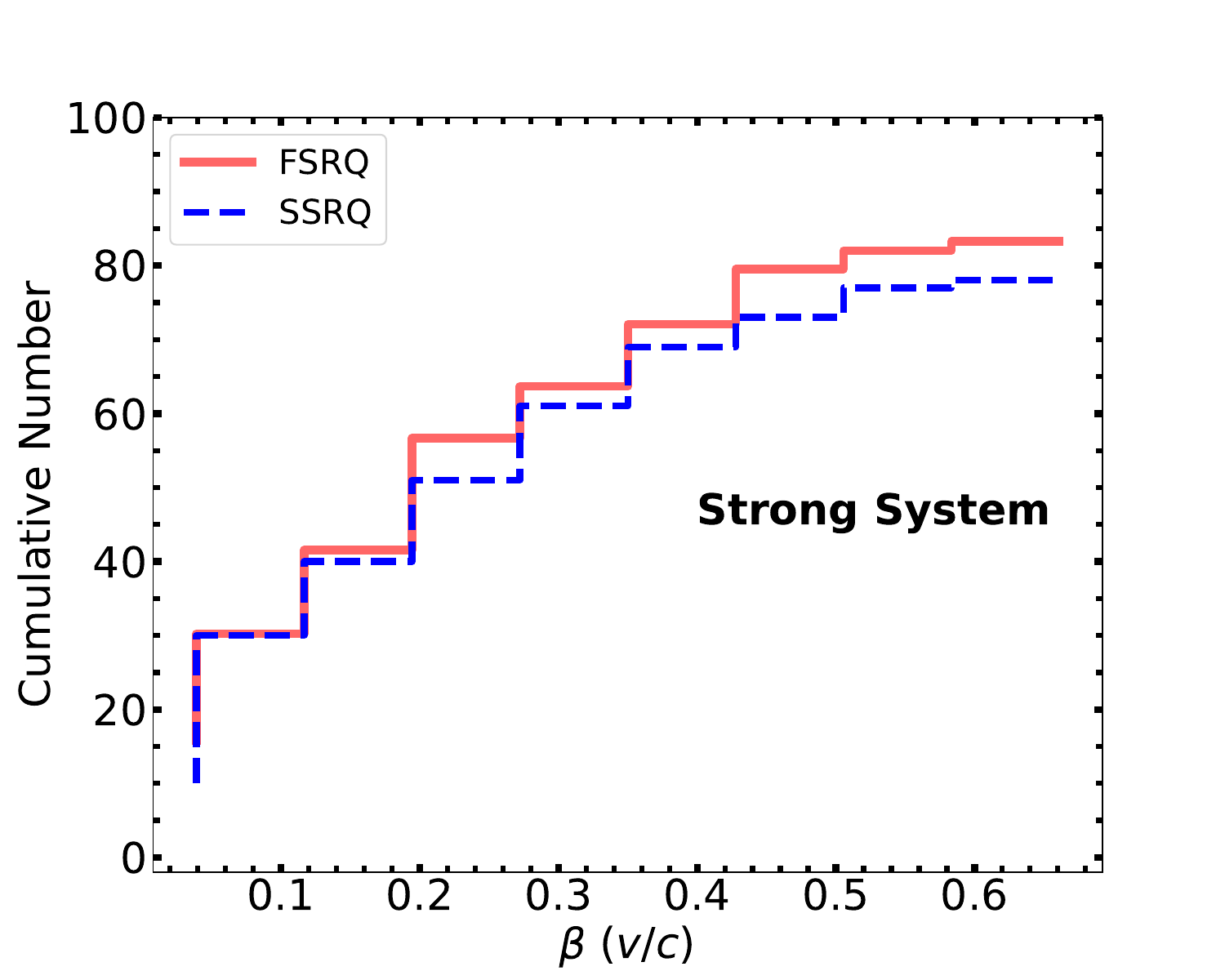}}
   \caption{Cumulative number for weak (top panel) and strong (bottom panel) \mgii~absorption systems as a function of $\beta$ detected towards the FSRQs (solid red-line) and SSRQs (dashed blue-line) after excluding the systems with a velocity offset of $5000$~\kms. To ensure a comparable $\beta$-path length between FSRQ and SSRQ sightlines, we scaled the number of \mgii~absorbers along FSRQs within each $\beta$ bin by the ratio of the $\beta$-path length ($\Delta \beta$) of FSRQs to that of SSRQs (see text in Section~\ref{subsec:result_dndb}).}
   \label{fig: beta_cumulative}
\end{figure}

Similar to Section~\ref{subsec:result_dndz}, this section examines the incidence rate and the cumulative number distribution of weak and strong \mgii~absorbers per unit $\beta$-path interval, \dndb, as a function of $\beta$ along FSRQ and SSRQ sightlines. In Fig.~\ref{fig: dn_db} we shows the incidence rate of intervening \mgii\ absorbers as a function of $\beta$, averaged over $\beta$ bins of width 0.2, for strong (bottom panel) and weak (top panel) \mgii\ absorbers along FSRQ (red circles)  and SSRQ (blue triangles) sightlines.  As evident from the figure, the \dndb\ distributions for FSRQs and SSRQs are consistent with each other for both strong ($P_{\text{null}} = 0.99 $) and weak ($P_{\text{null}} = 0.60$) systems, indicating no statistically significant difference. Next, Fig.~\ref{fig: beta_cumulative} presents the cumulative number of strong (bottom panel) and weak (top panel)  \mgii\ absorbers along FSRQ (solid red) and SSRQ (dashed blue) sightlines as a function of $\beta$. As in Section~\ref{subsec:result_dndz}, to account for the excess $\beta$-path in FSRQs, we normalize number of \mgii\ counts in each $\beta$-bin for FSRQs by the ratio  $ \Delta \beta_{\text{FSRQ}}/\Delta \beta_{\text{SSRQ}}$
where $\Delta \beta_{\text{FSRQ}}$ and $\Delta \beta_{\text{SSRQ}}$ represent the $\beta$-path lengths in individual $ \beta$-bin for FSRQs and SSRQs, respectively. As can be seen from the lower panel of Fig.~\ref{fig: beta_cumulative} that the normalized cumulative number of strong \mgii\ absorbers is indistinguishable between FSRQs and SSRQs across all $\beta$ values. For weak systems, a mild difference is observed: the normalized cumulative number is marginally lower for FSRQs compared to SSRQs at lower $\beta$ values ($\beta < 0.3$). At higher $\beta$ values ($\beta > 0.3$), this trend reverses, with FSRQs exhibiting a marginally higher number of weak \mgii\ absorbers than SSRQs.
\begin{figure}
  \begin{overpic}[width=0.24\textwidth, height=0.19\textwidth]{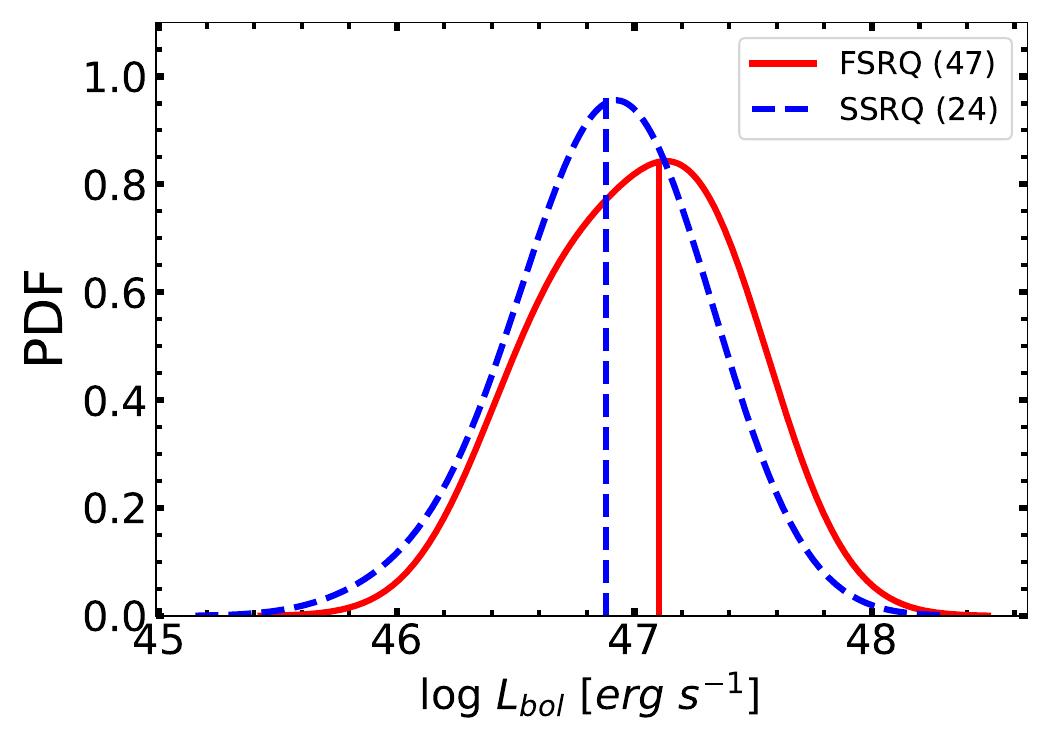}
    \put(18,60){\small $P_{\mathrm{null}} = 0.02$}
  \end{overpic}
  \begin{overpic}[width=0.24\textwidth, height=0.19\textwidth]{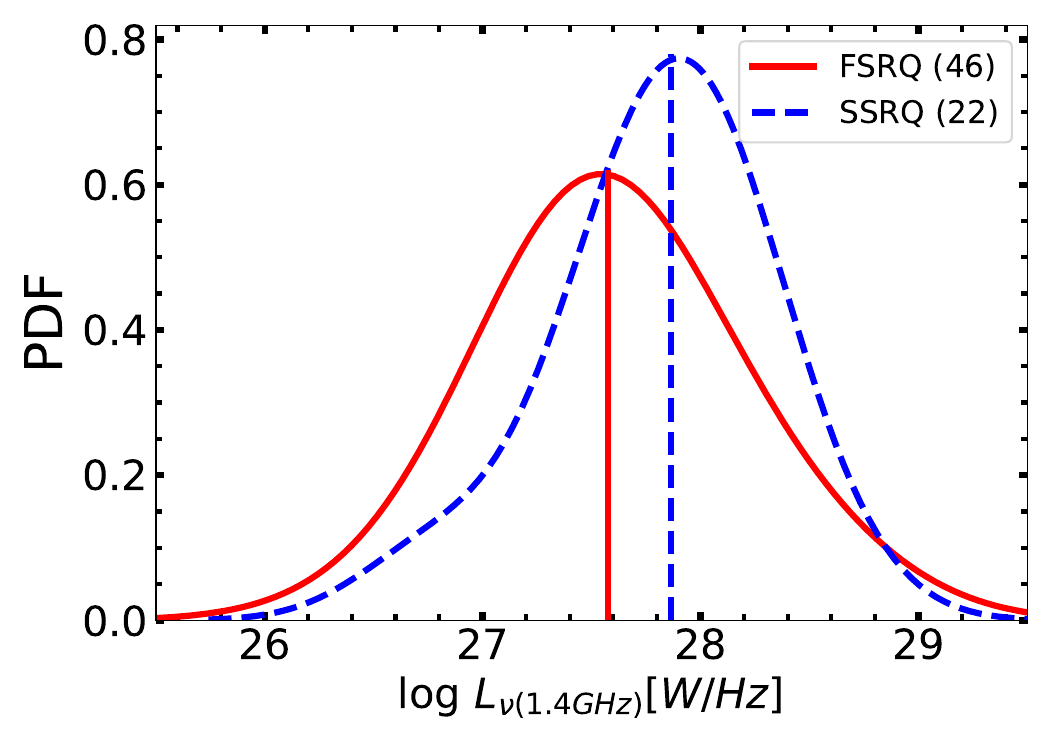}
    \put(18,60){\small $P_{\mathrm{null}} = 0.007$}
  \end{overpic}
  \includegraphics[width=0.24\textwidth, height=0.19\textwidth]{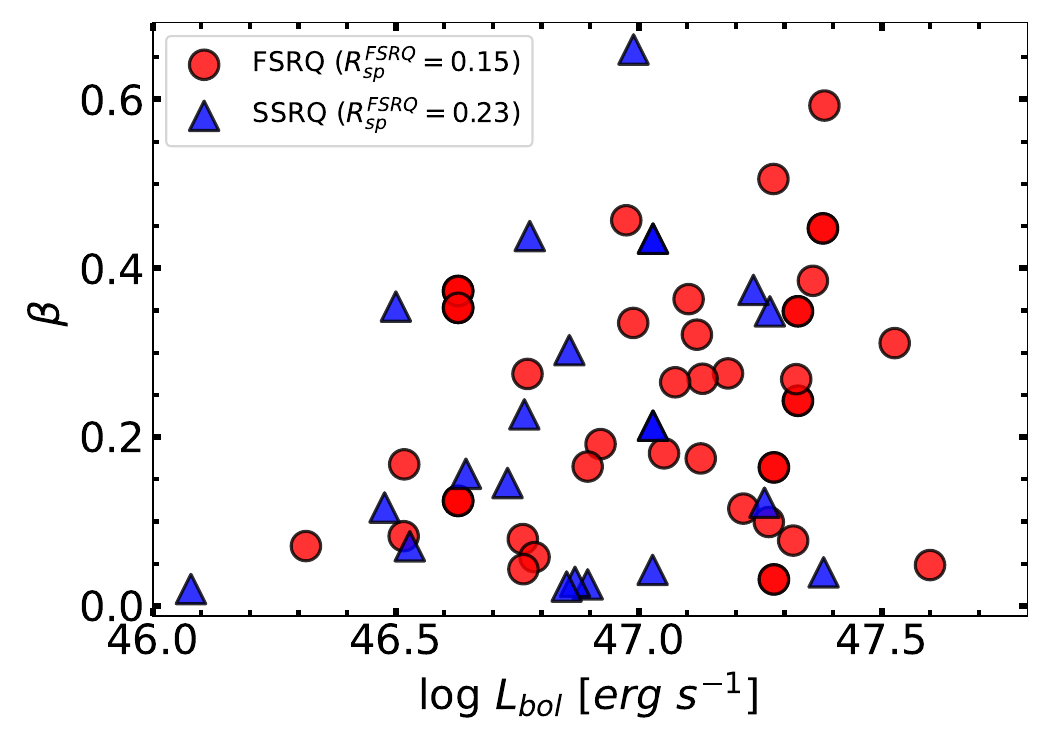}
  \includegraphics[width=0.25\textwidth, height=0.19\textwidth]{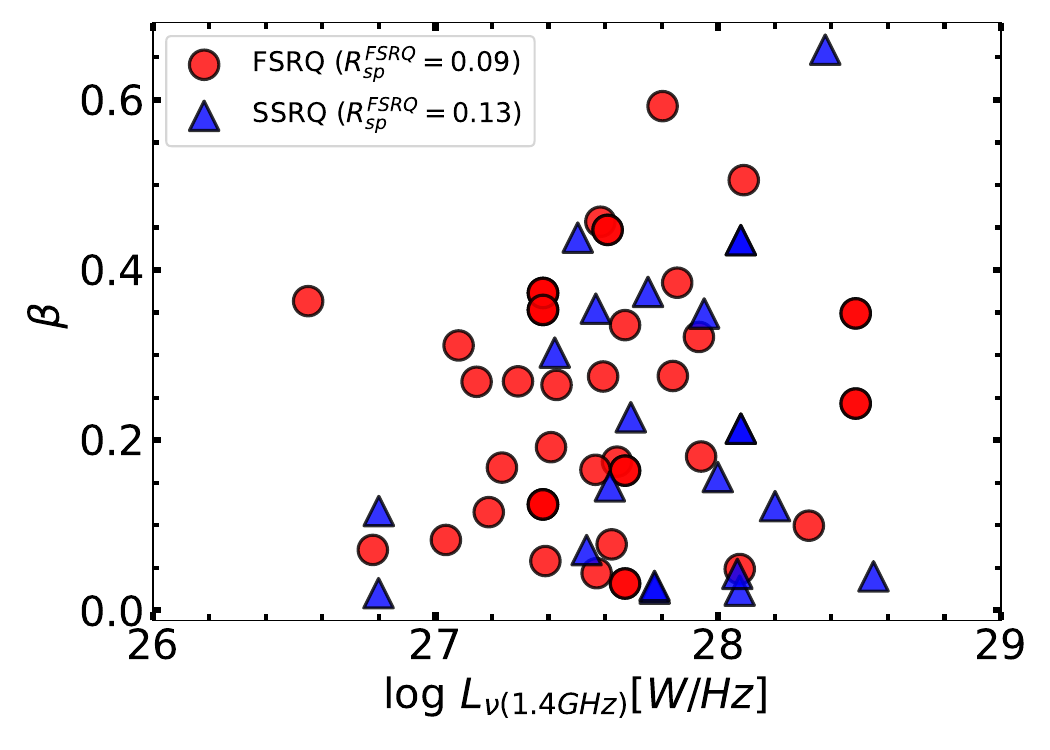}
  \caption{\emph{Top panels:} The Kernel Density Estimate (KDE) distributions plot of bolometric luminosity ($L_{\text {bol}}$) (left) and radio luminosity at 1.4 GHz ($L_{\text {1.4GHz}}$) (right) for the sample of FSRQs and SSRQs with detected weak absorption systems (total number given in parenthesis). \emph{Bottom panels:} The scatter plots of the velocity offset parameter $\beta$ versus bolometric luminosity (left) and radio luminosity $L_{\text {1.4GHz}}$) (right) for the same sample with the Spearman's rank correlation coefficient ($R_{\text{sp}}$) given in parentheses.}
  \label{fig: intrinsic_prop}
\end{figure}
\section{Discussion and Conclusions}
\label{discussion}
The incidence rate of \mgii~absorbers along the sightlines of various radio-loud quasar subclasses has been widely studied \citep[e.g., see][]{Bergeron2011A&A...525A..51B,Chand2012ApJ...754...38C,Joshi2013MNRAS.435..346J,mishra2018incidence}. However, direct comparisons between quasars with different jet orientations—such as FSRQs and SSRQs—remain limited, partly due to variations in sample selection and classification criteria. \par
In this study, we analyze the incidence rate, \dndz, of \mgii~absorption systems along the sightlines of 441 FSRQs (with $\alpha_{\text{radio}} > -0.3$) and 464 SSRQs ($\alpha_{\text{radio}} < -0.7$), representing one of the largest and well-characterized samples to date (see Section~\ref{sample}) based on the jet orientations. The classification of these two well-defined subclasses of radio-loud quasars—FSRQs and SSRQs—is based on the total intensity spectral index ($\alpha_{\text{radio}}$) derived from multi-frequency spectral energy distributions (SEDs) fitting, as described in FR14 (see Table~\ref{sample_table}). For both subclasses, we estimate the incidence rate of the weak ($0.3~\ang < W_{r}(2796) <1.0~\ang$, EW) and strong ($W_{r}(2796) \geq~1.0~\ang$, EW) \mgii~absorbers and present their \dndz~as a function of redshift (see Fig.~\ref{fig: dn_dz}), normalized cumulative distribution as a function of redshift (see Fig.~\ref{fig: cummulative_dz}), \dndb~as a function of $\beta$ (see Fig.~\ref{fig: dn_db}) and the normalized cumulative distribution as a function of $\beta$ (see Fig.~\ref{fig: beta_cumulative}). \par
We show in Fig.~\ref{fig: dn_dz} the \dndz~distributions as a function of redshift for weak and strong \mgii~systems. We find no significant difference in the \dndz~distributions of the FSRQ and SSRQ samples with redshift for weak and strong \mgii~systems (see Fig.~\ref{fig: dn_dz}). This trend holds across various subsamples of FSRQs and SSRQs, grouped according to their archival sources, as summarized in Table~\ref{dndz_result_table}. We note that in Table~\ref{dndz_result_table} the apparent excess in \dndz\ for weak \mgii\ absorbers toward non-SDSS SSRQs, relative to FSRQs, is based on a very limited redshift path ($\Delta z \approx$ 3.5 in contrast to $\Delta z \approx$ 68 for FSRQ), owing to the small number of non-SDSS SSRQ sightlines in this subgroup. 
Therefore, to draw a firm conclusion larger sample of  non-SDSS sightlines for SSRQs are also needed for a statistically robust analysis. Our findings align with those of \citet{mishra2018incidence}, who reported no excess in the incidence rate of intervening \mgii\ absorbers toward blazars compared to normal quasars. In Fig.~\ref{fig: cummulative_dz}, we present the cumulative distribution of weak and strong \mgii\ absorbers, normalized by redshift path, as a function of redshift for FSRQs and SSRQs. The distributions of strong absorbers appear similar between the two populations (KS-test P$_{null}$ of 0.73), whereas those of weak absorbers show marginal differences (KS-test P$_{null}$ of 0.35; significant at the 1$\sigma$ level), with FSRQs exhibiting a systematically higher number of intervening weak absorbers compared to SSRQs. A similar excess in the cumulative distribution of \mgii~absorbers is also observed in blazars compared to quasars, as reported by \citet{mishra2018incidence}, for intervening systems (velocity offset > 5000 \kms), including both weak and strong absorbers. This excess persists up to velocities of 0.2c, particularly for weak \mgii~systems. The authors attributed this to weak absorbers being accelerated outward by powerful blazar jets. This aligns with jet dynamics studies \citep[e.g.,][]{2017IAUS..324..141K}, which suggest that relativistic jets can remain globally stable against external disturbances, allowing weaker clumps to be more frequently accelerated without disrupting the jet. \par
To further investigate this scenario, we examine the \dndb\ evolution as a function of $\beta$, as shown in Fig.~\ref{fig: dn_db}. The \dndb\ distributions are statistically consistent between the FSRQ and SSRQ samples for both weak and strong \mgii\ absorbers with KS-test P$_{null}$ of 0.6 and 1.0, respectively. We also examine the cumulative number distribution as a function of $\beta$ for weak and strong \mgii\ absorbers in FSRQs and SSRQs, presented in Fig.~\ref{fig: beta_cumulative}. While the distribution for strong systems remains similar across both quasar populations, the weak absorber distribution shows a crossover near $\beta \sim 0.3$: FSRQs exhibit a slightly lower number of absorbers than SSRQs for $\beta \lesssim 0.3$, whereas for $\beta \gtrsim 0.3$, FSRQs display a marginally higher number of weak absorbers compared to SSRQs. \par
To investigate whether the redistribution of weak \mgii\ absorbers is linked to the intrinsic properties of quasars, we examine in Fig.~\ref{fig: intrinsic_prop} the distributions of bolometric luminosity ($L_{\text{bol}}$, top left) and non-thermal rest-frame radio luminosity at 1.4 GHz ($L_{1.4,\mathrm{GHz}}$, top right) using a non-parametric kernel density estimation \citep[KDE][]{    Silverman1986}, for our FSRQ and SSRQ samples. \par
The $L_{\text{bol}}$ values for 47 FSRQs and 24 SSRQs with weak \mgii\ absorbers are obtained from the SDSS DR16 quasar catalog by \citet{wu2022catalog}, while the $L_{1.4,\mathrm{GHz}}$ measurements for 46 FSRQs and 22 SSRQs with \mgii\ absorbers are obtained from the FR14 catalog. As can be seen in Fig.~\ref{fig: intrinsic_prop} that the $L_{\text{bol}}$ and $L_{1.4,\mathrm{GHz}}$ distribution among the FSRQs and SSRQs with weak \mgii\ absorbers are significantly different, with $P_{\text{null}}$ values of the KS-test being less than the 5 $\%$. We find that the FSRQs with weak \mgii\ absorbers are mildly skewed towards the higher $L_{\text{bol}}$ and lower $L_{1.4,\mathrm{GHz}}$ compared to the SSRQs. \\
We also assess in Fig.~\ref{fig: intrinsic_prop} (bottom two panels) how the $\beta$ values of the weak \mgii\ absorbers correlate with the $L_{\text{bol}}$ and the $L_{1.4,\mathrm{GHz}}$ of FSRQs and SSRQs, to investigate whether these absorbers are influenced by radiation-driven outflows or quasar jets \citep[see][]{Sharma2013}. The bottom-left panel of Fig.~\ref{fig: intrinsic_prop} presents the $\beta$ values of the weak \mgii\ absorbers for 47 FSRQs and 24 SSRQs as a function of $L_{\text{bol}}$. The Spearman correlation coefficients between $\beta$ and $L_{\text{bol}}$ are 0.14 for FSRQs and 0.23 for SSRQs, suggesting a weak correlation in both cases. Similarly, the bottom-right panel of  Fig.~\ref{fig: intrinsic_prop} shows the $\beta$ values for 46 FSRQs and 22 SSRQs, with detected weak \mgii~systems, as a function of $L_{1.4,\mathrm{GHz}}$. Here too, we find a very weak correlation, with Spearman coefficients of 0.09 for FSRQs and 0.13 for SSRQs. These results indicate that intervening \mgii\ absorbers show weak or negligible correlations with both $L_{\text{bol}}$ and $L_{1.4,\mathrm{GHz}}$. A similar study was conducted by \citet{Joshi2013MNRAS.435..346J}, who investigated the incidence of strong \mgii~absorbers toward CDQs and LDQs, based on their radio morphology. A detailed comparison with their results is warranted. In contrast to their classification scheme, our analysis distinguishes between FSRQs and SSRQs based on the spectral index ($\alpha$) derived from multi-wavelength SEDs. This classification more directly traces jet orientation, whereas radio morphology can be influenced by projection and resolution effects \citep{Urry1995}. A cross-match between our sample and that of \citet{Joshi2013MNRAS.435..346J} reveals no significant overlap, except for five common sources. Both studies apply a spectral SNR threshold of more than 5 for absorber identification and redshift path estimation, relying primarily on SDSS sightlines. However, \citet{Joshi2013MNRAS.435..346J} analyzes a significantly larger sample, with redshift paths of $\Delta z = 2333.01$ for CDQs and 922.11 for LDQs, compared to our redshift path coverage of 414.16 for FSRQs and 341.8 for SSRQs.  Their analysis is limited to strong \mgii~absorbers ($W_r^{2796} > 1$\AA), whereas our study includes both strong and weak systems, offering a more complete view of the absorber population. \citet{Joshi2013MNRAS.435..346J} report a statistically significant excess in the \dndz\ of strong \mgii~ absorbers toward CDQs relative to LDQs, with a factor of $\sim$1.13$\pm$0.10. In contrast, we find no such excess in either strong or weak absorber populations when comparing FSRQs and SSRQs. This discrepancy may arise from differences in sample size and redshift path coverage; our smaller dataset may lack the statistical power to detect subtle variations in \dndz. Alternatively, the difference may be physical. Our orientation-based classification more directly probes relativistic jet alignment, whereas CDQ/LDQ morphology may be affected by projection and resolution effects depending on luminosity and redshift, which may introduce additional scatter in the comparison. These explanations are not mutually exclusive and underscore the complexity of interpreting absorber statistics in radio-loud quasars. Future studies with larger, uniformly selected, jet-oriented samples will be crucial for disentangling these effects.
In summary, the presence of strong jets in quasars does not significantly affect the average \dndz\ distribution, as evidenced by the similar trends observed between FSRQs and SSRQs in our analysis (Fig.~\ref{fig: dn_dz}, Table~\ref{dndz_result_table}). This finding aligns with previous studies \citep{chand2012incidence, Joshi2013MNRAS.435..346J, mishra2018incidence}, which also report no notable variation in \dndz\ across different quasar types. However, for FSRQs—where the jets are closely aligned with the line of sight—\mgii\ absorbers, particularly the weak ones, tend to be redistributed toward higher velocities, consistent with the trends seen in the cumulative distributions (Fig.~\ref{fig: cummulative_dz}, Fig.~\ref{fig: beta_cumulative}) and as previously noted by \citet{mishra2018incidence}.
\section{Summary}
\label{Sect:summary}
In summary, we conclude that :
(i) The incidence rate, \dndz, of intervening \mgii~absorbers shows no significant excess for either strong or weak systems toward FSRQs compared to SSRQs. A marginal (non-significant) indication of excess in weak absorbers remains inconclusive, underscoring the need for a larger sample to clarify this trend. (ii) The \dndb~distribution of \mgii~absorbers along FSRQs and SSRQs are statistically similar, with a KS-test P$_{null}$ of 0.60 and 0.99 for weak and strong systems, respectively. (iii) The cumulative distribution of weak \mgii~absorbers (normalized in redshift path) is marginally lower for FSRQs compared to SSRQs at lower $\beta$ values ($\beta < 0.3$), but shows an excess at higher $\beta$. This suggests that, while the intrinsic abundance of \mgii~absorbers is similar along both FSRQ and SSRQ sightlines, the more aligned relativistic jets of FSRQs cluster weak absorbers at higher $\beta$ values, consistent with the AGN unification model and the jet dynamics studies allowing weaker clumps to be more frequently accelerated without disrupting the jet.

\section*{Acknowledgements}
We thank the anonymous referee for constructive comments and suggestions, which helped improve the clarity and quality of this manuscript. RK and HC are grateful to IUCAA for the hospitality and HPC facility under the IUCAA Associate Programme.

\section*{Data Availability}

The data used in this study are publicly available in the SDSS-DR16, SQUAD, and KODIAQ Data Releases.



\bibliographystyle{mnras}
\bibliography{references} 

\begin{thebibliography}{}
\makeatletter
\relax
\def\mn@urlcharsother{\let\do\@makeother \do\$\do\&\do\#\do\^\do\_\do\%\do\~}
\def\mn@doi{\begingroup\mn@urlcharsother \@ifnextchar [ {\mn@doi@} {\mn@doi@[]}}
\def\mn@doi@[#1]#2{\def\@tempa{#1}\ifx\@tempa\@empty \href {http://dx.doi.org/#2} {doi:#2}\else \href {http://dx.doi.org/#2} {#1}\fi \endgroup}
\def\mn@eprint#1#2{\mn@eprint@#1:#2::\@nil}
\def\mn@eprint@arXiv#1{\href {http://arxiv.org/abs/#1} {{\tt arXiv:#1}}}
\def\mn@eprint@dblp#1{\href {http://dblp.uni-trier.de/rec/bibtex/#1.xml} {dblp:#1}}
\def\mn@eprint@#1:#2:#3:#4\@nil{\def\@tempa {#1}\def\@tempb {#2}\def\@tempc {#3}\ifx \@tempc \@empty \let \@tempc \@tempb \let \@tempb \@tempa \fi \ifx \@tempb \@empty \def\@tempb {arXiv}\fi \@ifundefined {mn@eprint@\@tempb}{\@tempb:\@tempc}{\expandafter \expandafter \csname mn@eprint@\@tempb\endcsname \expandafter{\@tempc}}}

\bibitem[\protect\citeauthoryear{Aldcroft, Bechtold  \& Elvis}{Aldcroft et~al.}{1994}]{aldcroft1994mg}
Aldcroft T.~L.,  Bechtold J.,   Elvis M.,  1994, Astrophysical Journal Supplement Series (ISSN 0067-0049), vol. 93, no. 1, p. 1-46, 93, 1

\bibitem[\protect\citeauthoryear{{Anderson}, {Weymann}, {Foltz}  \& {Chaffee}}{{Anderson} et~al.}{1987}]{Anderson1987AJ.....94..278A}
{Anderson} S.~F.,  {Weymann} R.~J.,  {Foltz} C.~B.,   {Chaffee} Jr. F.~H.,  1987, \mn@doi [\aj] {10.1086/114468}, \href {http://adsabs.harvard.edu/abs/1987AJ.....94..278A} {94, 278}

\bibitem[\protect\citeauthoryear{{Bahcall} \& {Salpeter}}{{Bahcall} \& {Salpeter}}{1966}]{Bahcall1966ApJ...144..847B}
{Bahcall} J.~N.,  {Salpeter} E.~E.,  1966, \mn@doi [\apj] {10.1086/148675}, \href {http://adsabs.harvard.edu/abs/1966ApJ...144..847B} {144, 847}

\bibitem[\protect\citeauthoryear{Baker, Hunstead, Athreya, Barthel, de Silva, Lehnert  \& Saunders}{Baker et~al.}{2002}]{baker2002associated}
Baker J.~C.,  Hunstead R.~W.,  Athreya R.~M.,  Barthel P.~D.,  de Silva E.,  Lehnert M.~D.,   Saunders R.~D.,  2002, The Astrophysical Journal, 568, 592

\bibitem[\protect\citeauthoryear{Bergeron}{Bergeron}{1986}]{bergeron1986mg}
Bergeron J.,  1986, Astronomy and Astrophysics, 155, L8

\bibitem[\protect\citeauthoryear{Bergeron \& Boiss{\'e}}{Bergeron \& Boiss{\'e}}{1991}]{bergeron1991sample}
Bergeron J.,  Boiss{\'e} P.,  1991, {A\&A}, 243, 344

\bibitem[\protect\citeauthoryear{{Bergeron}, {Boiss{\'e}}  \& {M{\'e}nard}}{{Bergeron} et~al.}{2011}]{Bergeron2011A&A...525A..51B}
{Bergeron} J.,  {Boiss{\'e}} P.,   {M{\'e}nard} B.,  2011, \mn@doi [\aap] {10.1051/0004-6361/201015265}, \href {http://adsabs.harvard.edu/abs/2011A%26A...525A..51B} {525, A51}

\bibitem[\protect\citeauthoryear{Berk et~al.,}{Berk et~al.}{2008}]{berk2008average}
Berk D.~V.,  et~al., 2008, The Astrophysical Journal, 679, 239

\bibitem[\protect\citeauthoryear{{Chand} \& {Gopal-Krishna}}{{Chand} \& {Gopal-Krishna}}{2012}]{Chand2012ApJ...754...38C}
{Chand} H.,  {Gopal-Krishna} 2012, \mn@doi [\apj] {10.1088/0004-637X/754/1/38}, \href {http://adsabs.harvard.edu/abs/2012ApJ...754...38C} {754, 38}

\bibitem[\protect\citeauthoryear{Chand et~al.}{Chand et~al.}{2012}]{chand2012incidence}
Chand H.,  et~al., 2012, \apj, 754, 38

\bibitem[\protect\citeauthoryear{Chen \& Ho}{Chen \& Ho}{2025}]{chen2025detection}
Chen Z.-F.,  Ho L.~C.,  2025, The Astrophysical Journal, 980, 159

\bibitem[\protect\citeauthoryear{Chen, Qin, Cai, Zhou, Chen, Pang, Wang  \& Cheng}{Chen et~al.}{2023}]{chen2023study}
Chen Z.-F.,  Qin H.-C.,  Cai J.-T.,  Zhou Y.-T.,  Chen Z.-G.,  Pang T.-T.,  Wang Z.-W.,   Cheng K.-F.,  2023, The Astrophysical Journal Supplement Series, 265, 46

\bibitem[\protect\citeauthoryear{Christensen et~al.,}{Christensen et~al.}{2017}]{christensen2017solving}
Christensen L.,  et~al., 2017, Astronomy \& Astrophysics, 608, A84

\bibitem[\protect\citeauthoryear{Churchill, Rigby, Charlton  \& Vogt}{Churchill et~al.}{1999}]{churchill1999population}
Churchill C.~W.,  Rigby J.~R.,  Charlton J.~C.,   Vogt S.~S.,  1999, The Astrophysical Journal Supplement Series, 120, 51

\bibitem[\protect\citeauthoryear{Churchill, Kacprzak  \& Steidel}{Churchill et~al.}{2005}]{churchill2005mgii}
Churchill C.~W.,  Kacprzak G.~G.,   Steidel C.~C.,  2005, Proceedings of the International Astronomical Union, 1, 24

\bibitem[\protect\citeauthoryear{{Farnes}, {Gaensler}  \& {Carretti}}{{Farnes} et~al.}{2014}]{Farnes2014a}
{Farnes} J.~S.,  {Gaensler} B.~M.,   {Carretti} E.,  2014, \mn@doi [\apjs] {10.1088/0067-0049/212/1/15}, \href {https://ui.adsabs.harvard.edu/abs/2014ApJS..212...15F} {212, 15}

\bibitem[\protect\citeauthoryear{{Gehrels}}{{Gehrels}}{1986}]{Gehrels1986ApJ...303..336G}
{Gehrels} N.,  1986, \mn@doi [\apj] {10.1086/164079}, \href {http://adsabs.harvard.edu/abs/1986ApJ...303..336G} {303, 336}

\bibitem[\protect\citeauthoryear{{Joshi}, {Chand}  \& {Gopal-Krishna}}{{Joshi} et~al.}{2013}]{Joshi2013MNRAS.435..346J}
{Joshi} R.,  {Chand} H.,   {Gopal-Krishna} 2013, \mn@doi [\mnras] {10.1093/mnras/stt1294}, \href {http://adsabs.harvard.edu/abs/2013MNRAS.435..346J} {435, 346}

\bibitem[\protect\citeauthoryear{Kapahi, Athreya, Subrahmanya, Baker, Hunstead, McCarthy  \& van Breugel}{Kapahi et~al.}{1998}]{kapahi1998molonglo}
Kapahi V.~K.,  Athreya R.~M.,  Subrahmanya C.,  Baker J.~C.,  Hunstead R.~W.,  McCarthy P.~J.,   van Breugel W.,  1998, The Astrophysical Journal Supplement Series, 118, 327

\bibitem[\protect\citeauthoryear{{Komissarov}}{{Komissarov}}{2017}]{2017IAUS..324..141K}
{Komissarov} S.,  2017, in IAU Symposium. pp 141--148 (\mn@eprint {arXiv} {1702.06059}), \mn@doi{10.1017/S1743921317001922}

\bibitem[\protect\citeauthoryear{{Kulkarni}, {Meiring}, {Som}, {P{\'e}roux}, {York}, {Khare}  \& {Lauroesch}}{{Kulkarni} et~al.}{2012}]{Kulkarni2012ApJ...749..176K}
{Kulkarni} V.~P.,  {Meiring} J.,  {Som} D.,  {P{\'e}roux} C.,  {York} D.~G.,  {Khare} P.,   {Lauroesch} J.~T.,  2012, \mn@doi [\apj] {10.1088/0004-637X/749/2/176}, \href {http://adsabs.harvard.edu/abs/2012ApJ...749..176K} {749, 176}

\bibitem[\protect\citeauthoryear{{Mishra} \& {Muzahid}}{{Mishra} \& {Muzahid}}{2022}]{Mishra2022}
{Mishra} S.,  {Muzahid} S.,  2022, \mn@doi [\apj] {10.3847/1538-4357/ac7155}, \href {https://ui.adsabs.harvard.edu/abs/2022ApJ...933..229M} {933, 229}

\bibitem[\protect\citeauthoryear{Mishra, Chand, Krishna, Joshi, Shchekinov  \& Fatkhullin}{Mishra et~al.}{2018}]{mishra2018incidence}
Mishra S.,  Chand H.,  Krishna G.,  Joshi R.,  Shchekinov Y.,   Fatkhullin T.,  2018, \mnras, 473, 5154

\bibitem[\protect\citeauthoryear{Murphy, Kacprzak, Savorgnan  \& Carswell}{Murphy et~al.}{2019}]{murphy2019uves}
Murphy M.~T.,  Kacprzak G.~G.,  Savorgnan G.~A.,   Carswell R.~F.,  2019, \mnras, 482, 3458

\bibitem[\protect\citeauthoryear{{Nestor}, {Turnshek}  \& {Rao}}{{Nestor} et~al.}{2005}]{Nestor2005ApJ...628..637N}
{Nestor} D.~B.,  {Turnshek} D.~A.,   {Rao} S.~M.,  2005, \mn@doi [\apj] {10.1086/427547}, \href {http://adsabs.harvard.edu/abs/2005ApJ...628..637N} {628, 637}

\bibitem[\protect\citeauthoryear{O’Meara, Lehner, Howk  \& Prochaska}{O’Meara et~al.}{2020}]{o2020third}
O’Meara J.~M.,  Lehner N.,  Howk J.~C.,   Prochaska J.~X.,  2020, \apj, 161, 45

\bibitem[\protect\citeauthoryear{{Raghunathan}, {Clowes}, {Campusano}, {S{\"o}chting}, {Graham}  \& {Williger}}{{Raghunathan} et~al.}{2016}]{Srinivasan2016MNRAS.463.2640R}
{Raghunathan} S.,  {Clowes} R.~G.,  {Campusano} L.~E.,  {S{\"o}chting} I.~K.,  {Graham} M.~J.,   {Williger} G.~M.,  2016, \mn@doi [\mnras] {10.1093/mnras/stw2095}, \href {http://adsabs.harvard.edu/abs/2016MNRAS.463.2640R} {463, 2640}

\bibitem[\protect\citeauthoryear{{Sharma}, {Nath}  \& {Chand}}{{Sharma} et~al.}{2013}]{Sharma2013}
{Sharma} M.,  {Nath} B.~B.,   {Chand} H.,  2013, \mn@doi [\mnras] {10.1093/mnrasl/slt015}, \href {https://ui.adsabs.harvard.edu/abs/2013MNRAS.431L..93S} {431, L93}

\bibitem[\protect\citeauthoryear{Shen \& M{\'e}nard}{Shen \& M{\'e}nard}{2012}]{shen2012link}
Shen Y.,  M{\'e}nard B.,  2012, The Astrophysical Journal, 748, 131

\bibitem[\protect\citeauthoryear{{Silverman}}{{Silverman}}{1986}]{Silverman1986}
{Silverman} B.~W.,  1986, {Density estimation for statistics and data analysis}

\bibitem[\protect\citeauthoryear{Steidel}{Steidel}{1995}]{steidel1995nature}
Steidel C.~C.,  1995, in QSO Absorption Lines: Proceedings of the ESO Workshop Held at Garching, Germany, 21--24 November 1994. pp 139--152

\bibitem[\protect\citeauthoryear{Steidel \& Sargent}{Steidel \& Sargent}{1992}]{steidel1992mg}
Steidel C.~C.,  Sargent W.~L.,  1992, Astrophysical Journal Supplement Series (ISSN 0067-0049), vol. 80, no. 1, May 1992, p. 1-108. Research supported by NASA., 80, 1

\bibitem[\protect\citeauthoryear{{Stocke} \& {Rector}}{{Stocke} \& {Rector}}{1997}]{Stocke1997}
{Stocke} J.~T.,  {Rector} T.~A.,  1997, \mn@doi [\apjl] {10.1086/310962}, \href {https://ui.adsabs.harvard.edu/abs/1997ApJ...489L..17S} {489, L17}

\bibitem[\protect\citeauthoryear{{Tejos}, {Lopez}, {Prochaska}, {Bloom}, {Chen}, {Dessauges-Zavadsky}  \& {Maureira}}{{Tejos} et~al.}{2009}]{Tejos2009ApJ...706.1309T}
{Tejos} N.,  {Lopez} S.,  {Prochaska} J.~X.,  {Bloom} J.~S.,  {Chen} H.-W.,  {Dessauges-Zavadsky} M.,   {Maureira} M.~J.,  2009, \mn@doi [\apj] {10.1088/0004-637X/706/2/1309}, \href {http://adsabs.harvard.edu/abs/2009ApJ...706.1309T} {706, 1309}

\bibitem[\protect\citeauthoryear{{Urry} \& {Padovani}}{{Urry} \& {Padovani}}{1995}]{Urry1995}
{Urry} C.~M.,  {Padovani} P.,  1995, \mn@doi [\pasp] {10.1086/133630}, \href {https://ui.adsabs.harvard.edu/abs/1995PASP..107..803U} {107, 803}

\bibitem[\protect\citeauthoryear{{Wolfe}, {Gawiser}  \& {Prochaska}}{{Wolfe} et~al.}{2005}]{Wolfe2005ARA&A..43..861W}
{Wolfe} A.~M.,  {Gawiser} E.,   {Prochaska} J.~X.,  2005, \mn@doi [\araa] {10.1146/annurev.astro.42.053102.133950}, \href {http://adsabs.harvard.edu/abs/2005ARA%26A..43..861W} {43, 861}

\bibitem[\protect\citeauthoryear{Wu \& Shen}{Wu \& Shen}{2022}]{wu2022catalog}
Wu Q.,  Shen Y.,  2022, \apjs, 263, 42

\bibitem[\protect\citeauthoryear{{Zhu} \& {M{\'e}nard}}{{Zhu} \& {M{\'e}nard}}{2013}]{Zhu2013ApJ...770..130Z}
{Zhu} G.,  {M{\'e}nard} B.,  2013, \mn@doi [\apj] {10.1088/0004-637X/770/2/130}, \href {http://adsabs.harvard.edu/abs/2013ApJ...770..130Z} {770, 130}

\makeatother
\end{thebibliography}
\bsp	
\label{lastpage}
\end{document}